%% file: main.tex
\documentclass[a4paper,twocolumn,11pt,unpublished]{quantumarticle}
\pdfoutput=1
\usepackage[utf8]{inputenc}
\usepackage[english]{babel}
\usepackage[T1]{fontenc}
\usepackage{amsmath,amssymb,bm,amsthm,graphicx,booktabs}
\usepackage{microtype}
\usepackage[numbers,sort&compress]{natbib}
\usepackage{hyperref}
\usepackage{xstring,xcolor}
\let\originalbiblink\href
\newcommand{\canonicalbiblink}[2]{%
  \IfBeginWith{#1}{https://dx.doi.org/}{%
    \StrBehind{#1}{https://dx.doi.org/}[\bibdoisuffix]%
    \originalbiblink{https://doi.org/\bibdoisuffix}{#2}%
  }{\originalbiblink{#1}{#2}}}
\definecolor{linkink}{HTML}{234A71}
\hypersetup{colorlinks,allcolors=linkink}
\newtheorem{theorem}{Theorem}
\newtheorem{corollary}{Corollary}
\newtheorem{proposition}{Proposition}
\begin{document}
\title{Low-energy spectral-window truncation for compact \texorpdfstring{$U(1)$}{U(1)} lattice gauge theory}
\author{Yi Liang}
\orcid{0009-0006-3168-972X}
\email{iantrans2042@gmail.com}
\affiliation{Independent Researcher, Beijing, China}
\author{Chen Qian}
\orcid{0000-0002-3851-7713}
\email{qianchen@baqis.ac.cn}
\affiliation{Beijing Academy of Quantum Information Sciences, Beijing, China}
\date{}
\maketitle
\hypersetup{pdfauthor={Yi Liang and Chen Qian},pdftitle={Low-energy spectral-window truncation for compact U(1) lattice gauge theory},pdfcreator={},pdfproducer={}}
\begin{abstract}
\input{abstract}
\end{abstract}
\raggedbottom
\section{Introduction}
\input{sections/introduction}
\input{sections/setup}
\input{sections/tail}
\input{sections/spectroscopy}
\input{sections/dynamics}
\input{sections/coherent-example}
\input{sections/electric-square}
\input{sections/resources}
\input{sections/discussion}
\paragraph*{Author contributions.} All authors contributed equally to this work. During the preparation of this manuscript, the authors used large language models for proofreading, language editing, bibliographic searching, and code writing. The authors subsequently reviewed and edited the complete manuscript and take full responsibility for its final content.
\input{data-availability}
\appendix
\numberwithin{equation}{section}
\renewcommand{\theequation}{\thesection\arabic{equation}}
\input{sections/auxiliary}
\input{sections/coherent-proof}
\input{sections/regulator-details}
\input{sections/weighted-proof}
\bibliographystyle{quantum}
\begingroup\let\href\canonicalbiblink
\bibliography{refs}
\endgroup
\end{document}

%% file: abstract.tex
How many flux levels per link are needed to approximate low-energy states or their energies? We consider pure compact $U(1)$ gauge theory on a three-dimensional periodic cubic lattice at zero vacuum angle, in the gauge-invariant zero-global-flux sector. Fix the finite lattice and its spacing, the positive bare coupling, and the window width above the interacting vacuum. As the vector-norm tolerance $\epsilon$ tends to zero, the least cutoff for uniform approximation of window states in the natural hard-wall embedding satisfies $\Lambda\sim\log(1/\epsilon)/[2\log\log(1/\epsilon)]$. With energies measured in units of the finite-volume vacuum gap, the least cutoff for approximating all ordered window levels to the same absolute tolerance is asymptotically half as large. In both cases the interacting vacuum supplies the matching lower bound, while the upper bound holds on the entire original window. At weak coupling, a modified semiclassical weight and the energy cost of compensating flux reduce the sufficient cutoff for time evolution. In a strong-coupling example, seven flux levels per link certify an absolute truncation error below $3.12\times10^{-15}$ in inverse box-length units for each of 163 ordered levels and preserve an order-one change in the plaquette expectation of a coherent wave packet.

%% file: sections/introduction.tex
Putting space on a grid leaves an infinite local state space for the compact $U(1)$ gauge field~\cite{Wilson1974,Kogut1979,KogutSusskind1975}. The electric flux $E_l$ on link $l$ takes every integer value; its eigenstates form a basis of the rotor space $\ell^2(\mathbb Z)$. A finite quantum register~\cite{Feynman1982,Lloyd1996,JordanLeePreskill2012} cannot encode all these orthogonal states exactly. The electric hard wall~\cite{ByrnesYamamoto2006} retains the fluxes $-\Lambda,\ldots,\Lambda$ and deletes only hops leaving this interval; the matrix elements between retained fluxes are unchanged. This gives two accuracy tests for the same Hamiltonian: how closely the register can represent the original low-energy states, and how far the wall displaces their ordered energies. Fang and Tong~\cite{FangTong2026} give a sufficient register cutoff for the whole low-energy window even at weak coupling. How far could the cutoff be lowered before one of these two accuracy tests must fail?

We study pure compact $U(1)$ theory in three spatial dimensions, on a periodic cubic lattice with $N\ge3$ sites per direction, at vacuum angle $\theta=0$, in the sector with zero charge at every site and zero net flux through each coordinate plane. Fix the lattice and its spacing, the bare coupling $g>0$, and the window width $E_*\ge0$. The window projector $P_*$ selects the spectral interval $[E_0,E_0+E_*]$ in this sector, where $E_0$ is the untruncated vacuum energy. Write $u=\log(1/\epsilon)$ as the tolerance $\epsilon$ tends to zero. In the natural embedding of the hard-wall space, the least cutoff for representing every normalized window state within vector-norm error $\epsilon$ is asymptotic to $u/(2\log u)$. For the ordered hard-wall energies to approximate all original window levels, counting multiplicities, the least cutoff is instead asymptotic to $u/(4\log u)$ at absolute tolerance $\epsilon\Delta_{\rm vac}$, where $\Delta_{\rm vac}$ is the positive vacuum gap in this sector. Neither result requires an isolation gap at the upper window edge.

The same wave packet is a target for every choice of wall: it is the spectrum of the untruncated Hamiltonian, not a fresh selection of levels inside each finite register, that defines $P_*$. This matters for coherent wave packets whose relative phases evolve in time. Merely bounding their mean excitation energy would also admit inputs with quite different flux tails. An electric loop carrying $M$ units around an elementary lattice square (a plaquette) costs an energy of order $M^2$. For every $E_*>0$, giving such a loop probability of order $M^{-2}$ in a mixture with the vacuum can keep the mean excitation exactly at $E_*$. The probability of finding $|E_l|>M-1$ on a link of the loop is then bounded below by a constant times $M^{-2}$; Appendix~\ref{app:auxiliary} gives pure-state examples as well. Even with genuine window support, a bound $r$ on the tail norm of each vector of an orthonormal eigenbasis for a $D$-dimensional window gives a sum estimate of $\sqrt D\,r$ for a normalized superposition. Instead we estimate $\|\mathbf1_{\{|E_l|>\Lambda\}}P_*\|$, the largest discarded amplitude on link $l$ for a unit vector anywhere in the original window.

Tong et al.~\cite{Arxiv2110_06942} bound the dynamics for bounded initial local quantum numbers; their static estimate concerns an isolated nondegenerate eigenstate with finite first absolute moment. Other spectral estimates assume bounded sums of interaction strengths per site~\cite{Arxiv1406_3898} or use the window of an operator that has already been truncated~\cite[Supplementary Theorem~5]{Arxiv2411_02157}. By contrast, the window studied by Fang and Tong is a window of the original gauge theory. To complement a sufficient bound one needs an input allowed by that very promise on which truncation necessarily loses information. The witness must belong to the interacting model.

The upper bound is local, even though the window is measured from an extensive vacuum energy. Following the idea of a subsystem comparison~\cite{Arxiv2203_16012}, we remove the kinetic term on the chosen link and the four plaquette terms that meet it. The ground energy of the remaining lattice supplies a common reference for the window and the high-flux configurations. Gauss's law forces flux on the link to return through the rest of the lattice; minimizing the electric cost of this flow in the zero-winding sector raises the penalty for a large link flux. Only hops around incident plaquettes can change that flux, and each hop changes it by one unit. An operator equation between the high-flux and window subspaces (a Sylvester equation) then bounds each successive tail by the preceding tail times a factor of order $j^{-2}$ for large shell index $j$. Multiplying these factors gives factorial-squared suppression of the tail norm. For the converse, consider the configurations obtained by adding flux repeatedly around a single plaquette. In the electric basis the interacting vacuum amplitudes can be chosen nonnegative, so keeping just this path of successive hops gives a lower bound rather than a contribution that could be cancelled by another path. It has the same leading logarithm as the upper estimate.

The two tasks differ at the wall. In the Rayleigh--Ritz comparison, a hop across the wall connects the last retained flux shell $\Lambda$ to the first excluded shell $\Lambda+1$. After accounting for the normalization of the projected trial subspace, the energy-shift bound contains the product of two small shell amplitudes, whereas the driving term for real-time evolution has a window state on only one side of the boundary. Factorial suppression also appears in high-field analyses~\cite{Arxiv2508_00061,Arxiv2502_03533}, adjacent boundary coefficients enter Mathieu eigenvalue estimates~\cite{VolkmerMathieuRitz}, and quadratic sensitivity of Ritz values to subspace angles is well established~\cite{KnyazevArgentatiRitz}. Here the interacting Hamiltonian supplies a converse: its ground energy inside the wall exceeds the exact vacuum energy by at least the vacuum gap times the omitted vacuum probability. That single level fixes the leading rate for uniform absolute-energy accuracy.

Weak coupling poses a separate question at fixed evolution time and accuracy. We retain Fang and Tong's local vacuum comparison and weighted estimate on the entire window. On each shell where the direct Fang--Tong v1 bound is at most one, the modified weight gives a strictly smaller bound without using the return-flow estimate. Choosing the weight parameter to account for the compensating flux tightens it further. Fix the lattice and box size, the window width, a positive evolution time and a state-vector tolerance $0<\epsilon<1$. As $g\to0$, the ratio of the first cutoffs certified directly by the return-flow weight and by Fang--Tong v1, respectively, tends to $(2c_N)^{-1/4}$. Here $c_N=3N^3/[2(N^3-1)]$ is the minimum total squared flux of a real, divergence-free, zero-winding flow carrying unit flux on the chosen link. The improvement is in the leading coefficient, not the weak-coupling power.

The strong-coupling example in Sec.~\ref{sec:coherent-example} serves a different purpose. A window identified analytically contains the vacuum and a complete excited band. The same finite register bounds the errors in its ordered energies and preserves an order-one change in the plaquette expectation of a coherent wave packet. The polar part of an injective wall projection gives one isometric encoding for all window inputs, including those entangled with a reference. For a separate, prescribed superposition of flux configurations, two local four-link unitaries suffice for preparation without an interacting eigenvector.

Section~\ref{sec:setup} specifies the hard wall and the window. Sections~\ref{sec:tail} and~\ref{sec:spectroscopy} derive the tail and energy bounds and show how certified finite-wall energies can test membership in the original window. Sections~\ref{sec:dynamics}--\ref{sec:electric-square} treat evolution, encoding, preparation and readout. Section~\ref{sec:resources} gives register budgets for increasing precision, decreasing coupling and lattice refinement.

%% file: sections/setup.tex
\section{Regulator and spectral-window guarantee}
\label{sec:setup}

We consider pure compact $U(1)$ gauge theory formulated on a periodic cubic lattice with physical side length $\ell$, lattice spacing $a=\ell/N$, and linear size $N\geq3$. Under the Kogut--Susskind Hamiltonian framework~\cite{KogutSusskind1975}, an infinite-dimensional rotor Hilbert space $\ell^2(\mathbb Z)$ is assigned to every oriented link $l$. In the electric flux basis, the canonical electric field and link phase operators act according to
\begin{equation}
 E_l|n\rangle=n|n\rangle,\qquad U_l|n\rangle=|n+1\rangle,
 \qquad n\in\mathbb Z.
 \label{eq:rotor}
\end{equation}
The geometry contains $n_l=3N^3$ links, $n_p=3N^3$ elementary square plaquettes, and exactly $m=4$ plaquettes sharing any specified link. The requirement $N\geq3$ avoids self-intersecting plaquettes and small-torus identifications. For each oriented plaquette $p$, the magnetic holonomy is $U_p=\prod_{l\in\partial p}U_l^{c_{pl}}$ with $c_{pl}=+1$ if the link orientation agrees with the oriented boundary of $p$ and $c_{pl}=-1$ otherwise, and the gauge-invariant magnetic term is $C_p=(U_p+U_p^\dagger)/2$. In the absence of dynamical matter and at vanishing $\theta$ angle, the Hamiltonian at bare coupling $g>0$ is
\begin{equation}
 \begin{aligned}
  &H_a=\kappa_a\sum_l E_l^2+J_a\sum_p(1-C_p), \\
  &\kappa_a=\frac{g^2}{2a},\quad J_a=\frac{1}{g^2a}.
 \end{aligned}
 \label{eq:hamiltonian}
\end{equation}
Because the magnetic potential is bounded and smooth over the angle configuration torus, $H_a$ is self-adjoint on the domain of the unperturbed electric kinetic operator $\sum_l E_l^2$, is bounded below by zero, and possesses a purely discrete spectrum with compact resolvent at each finite lattice size.

We impose local Gauss constraints and select the sector with zero net electric flux across every coordinate plane. Write $x$ for a site modulo $N$ in each coordinate and $e_i$ for the positive unit vector in direction $i=1,2,3$. The electric operator $E_{x,i}$ on the link from $x$ to $x+e_i$ has eigenvalue $n_{x,i}\in\mathbb Z$ on a flux configuration $n$. Outgoing minus incoming flux gives the local charge; the signed flux across a coordinate cut gives the global electric flux in its normal direction. The corresponding commuting operators are
\begin{equation}
 G_x=\sum_{i=1}^3(E_{x,i}-E_{x-e_i,i}),\qquad
 F_i=\sum_{x:x_i=0}E_{x,i},
 \label{eq:constraints}
\end{equation}
whose joint kernel defines the physical subspace projector $P_{\rm phys,0}$. Plaquette hops preserve this sector. As established in Sec.~\ref{sec:tail}, the unique ground state of the complete Hamiltonian resides in this sector; we denote its vacuum energy by $E_0$. For any energy cutoff $E_*\geq0$, we set
\begin{equation}
 P_*=\mathbf1_{[E_0,E_0+E_*]}(H_a)P_{\rm phys,0}.
 \label{eq:window}
\end{equation}
This formulation accommodates arbitrary spectral degeneracies and requires no isolation gap at the window boundary $E_0+E_*$. When $E_*=0$, this is the vacuum projector. For $E_*>0$, support in this window is strictly stronger than a bound of $E_*$ on the mean excitation energy.

We regularize the unbounded rotor variables by imposing an integer electric flux cutoff $\Lambda\geq0$ on each link, following the endpoint-zero shift prescription of Byrnes and Yamamoto~\cite{ByrnesYamamoto2006}. The corresponding hard-wall projection and truncated Hamiltonian are
\begin{equation}
 \begin{aligned}
  &P_\Lambda=\prod_l\mathbf1_{[-\Lambda,\Lambda]}(E_l), \\
  &H_{a,\Lambda}=P_\Lambda H_aP_\Lambda\big|_{\operatorname{Ran}P_\Lambda}.
 \end{aligned}
 \label{eq:wall}
\end{equation}
Because $P_\Lambda$ is diagonal in the electric charge basis, it commutes with both $G_x$ and $F_i$, preserving the physical Gauss and flux sectors. Truncated link Hilbert spaces play a central role across digital quantum algorithms, analog quantum simulators, and tensor network algorithms~\cite{DalmonteMontangero2016,BanulsEtAl2020,ZoharCiracReznik2016,Zohar2022}, with the local qudit dimension increasingly tunable in laboratory setups~\cite{QuditLGT2025}. The cutoff analyzed here operates as a representation truncation~\cite{ZoharBurrello2015}: while it strictly preserves gauge covariance, the flux-shifting ladder operators cease to be unitary at the truncation boundaries. In this formulation, link raising and lowering operators retain unit interior matrix elements and simply truncate hops beyond $\pm\Lambda$. This structure differs from cyclic discrete-group shifts that wrap boundary hops back to the opposite endpoint, and from quantum link models whose spin-algebra generators alter the interior matrix elements~\cite{ChandrasekharanWiese1997}. While hard-wall dynamics can also be extracted by projecting larger cyclic group registers~\cite{Haase2021}, our results apply directly to the compression defined in Eq.~\eqref{eq:wall}.

Any electric flux concentrated on an individual link must be balanced across the remaining lattice to satisfy both the local Gauss law and the global flux-neutrality constraints. Within the neutral sector, the minimal squared $\ell_2$ norm of any real divergence-free flow carrying flux $k$ on that link is $c_N k^2$. For each shell index $j\in\mathbb Z_{\ge0}$, define
\begin{equation}
 \begin{aligned}
 c_N&=\frac{3N^3}{2(N^3-1)}, & A_g&=\frac{8}{g^4},\\
 s_a&=\frac{2aE_*}{g^2}, & d_j&=c_N(j+1)^2-s_a .
 \end{aligned}
 \label{eq:tail-parameters}
\end{equation}
Combining the kinetic penalty of this circulating return flow with an operator-norm bound on transitions into high-flux configurations yields the cumulative product certificate
\begin{equation}
 \begin{aligned}
 R_\Lambda&=\prod_{j=0}^{\Lambda} f_j,\qquad\\
 f_j&=\begin{cases}
 1,&d_j\le A_g,\\
 A_g/[2(d_j-A_g)],&j\ge1,\ d_j>2A_g,\\
 A_g/d_j,&\text{otherwise}.
 \end{cases}
 \end{aligned}
 \label{eq:tail-product}
\end{equation}
The first strictly subunit factor in this product occurs at the shell index $j_0=\lfloor\sqrt{(A_g+s_a)/c_N}\rfloor$.

\begin{theorem}[Whole-window truncation and dynamics]
\label{thm:main}
Let $N\geq3$, coupling $g>0$, window width $E_*\geq0$, and cutoff $\Lambda\in\mathbb Z_{\geq0}$. On each link $l$, the untruncated low-energy spectral projector $P_*$ obeys the local tail bound
\begin{equation}
 \big\|\mathbf1_{\{|E_l|>\Lambda\}}P_*\big\|\leq R_\Lambda.
 \label{eq:local-tail}
\end{equation}
This estimate applies uniformly to every normalized coherent superposition in $\operatorname{Ran}P_*$, with no prefactor depending on the dimension of that subspace. It assumes neither eigenvalue nondegeneracy nor an isolation gap at the window boundary $E_0+E_*$. Provided the cutoff satisfies $\sqrt{3N^3}R_\Lambda<1$, the normalized hard-wall projection $\psi_\Lambda=P_\Lambda\psi/\|P_\Lambda\psi\|$ is well defined for any such state $\psi$, and for all finite $T\geq0$ the original and compressed Schrödinger trajectories satisfy
\begin{equation}
 \begin{aligned}
  &\sup_{|t|\leq T}\big\|e^{-itH_a}\psi-e^{-itH_{a,\Lambda}}\psi_\Lambda\big\| \\
  &\leq B_NR_\Lambda, \\
  &B_N=2\sqrt{3N^3}+\frac{6TN^4}{g^2\ell}.
 \end{aligned}
 \label{eq:trajectory}
\end{equation}
Identifying the truncated state space with its natural included subspace of the original Hilbert space, let $O_\Lambda=(P_\Lambda OP_\Lambda)|_{\operatorname{Ran}P_\Lambda}$ for any bounded operator $O$ on the original space with $\|O\|\leq1$. Under the same conditions, the expectation of $O$ along the original evolution and that of $O_\Lambda$ along the truncated evolution differ by at most $2B_NR_\Lambda$ uniformly for $|t|\leq T$.
\end{theorem}

For fixed physical side length $\ell$, bare coupling $g>0$, spectral window cutoff $E_*$, and duration $T$, the tail bound obeys the asymptotic scaling $\log R_\Lambda=-2\Lambda\log\Lambda+O(\Lambda)$ uniformly in $N$. Hence, selecting $\Lambda_N=\lceil\beta\log N/\log\log N\rceil$ with any fixed $\beta>2$ suffices to drive the dynamical tracking error to zero in the refinement limit $N\to\infty$. Section~\ref{sec:tail} establishes the static local-flux estimate, Sec.~\ref{sec:dynamics} proves the dynamical tracking bound, and Sec.~\ref{sec:resources} evaluates sufficient cutoffs and corresponding local register sizes.

An unknown input supported in $\operatorname{Ran}P_*$ can be encoded by a single linear isometry rather than by normalizing its projection separately for each state. Section~\ref{sec:dynamics} constructs it whenever $\sqrt{n_l}R_\Lambda<1$. When the truncated hard-wall output is embedded into the ambient Hilbert space, the resulting channel error is bounded by $2B_NR_\Lambda$ in the unhalved diamond norm, even in the presence of arbitrary entanglement with an external reference system. A separate converse concerns the static hard-wall tail: at fixed lattice size, approximating the vacuum in state-vector norm already requires the leading cutoff scale.

%% file: sections/tail.tex
\section{Local energy separation and the spectral tail}
\label{sec:tail}

To control high electric flux on any chosen link $l$, one can decouple that link from the remainder of the lattice rather than attempting to track the extensive ground-state energy directly. Excising link $l$ entails dropping its kinetic term $\kappa_a E_l^2$ together with the $m=4$ plaquette terms that touch it. On the configuration space of the remaining $n_l-1$ link phases, this leaves the environment Hamiltonian
\begin{equation}
 \begin{aligned}
  &H_{\rm env}=\kappa_a\sum_{l'\ne l}E_{l'}^2 +J_a\sum_{p\not\ni l}(1-C_p), \\
  &e_{\rm env}=\inf\operatorname{spec}H_{\rm env}.
 \end{aligned}
 \label{eq:environment}
\end{equation}
The remaining link angles form the connected compact flat torus $\mathbb{T}^{n_l-1}$, on which $H_{\rm env}$ has an elliptic kinetic term and a smooth, real, bounded magnetic potential. The strictly positive free heat kernel and the bounded-potential Feynman--Kac representation~\cite{BaerPfaeffle2011FeynmanKac} give $e^{-tH_{\rm env}}$ a strictly positive integral kernel for every $t>0$. Compact resolvent supplies an attained lowest eigenvalue; the standard positivity-improving ground-state theorem makes $e_{\rm env}$ simple with an everywhere strictly positive normalized eigenfunction $\phi_{\rm env}$~\cite{KellerEtAl2015HeatKernels}. Every surviving plaquette avoids the deleted link, so its boundary has zero restricted divergence at each vertex and zero signed intersection with the original global electric-flux cuts. The restricted gauge and electric-cut symmetries commute with $H_{\rm env}$ and act by translations of the remaining angles, preserving positive real functions. Simplicity makes their action on $\phi_{\rm env}$ a scalar phase, and positivity and normalization fix this phase to one. Thus these symmetries act trivially on $\phi_{\rm env}$. Tensoring it with $|E_l=0\rangle$ yields an admissible normalized trial state in the original physical neutral sector.

In this composite trial state, every discarded plaquette $C_p$ contains either $U_l$ or $U_l^\dagger$ and thus has vanishing expectation value against $|E_l=0\rangle$. The kinetic contribution $\kappa_a E_l^2$ vanishes as well. Each deleted plaquette therefore contributes only its constant energy offset $J_a$, so the Rayleigh--Ritz variational principle provides the upper bound
\begin{equation}
 E_0\leq e_{\rm env}+mJ_a.
 \label{eq:vacuum-cost}
\end{equation}
Repeating the same heat-kernel positivity and Perron--Frobenius argument for the full Hamiltonian $H_a$ across all $n_l$ links confirms that the absolute vacuum is unique, strictly positive on the full torus, and invariant under all gauge and flux translations, so its energy is indeed the physical reference $E_0$ introduced in Sec.~\ref{sec:setup}. Subsystem environment isolation of this type has established antecedents in Hamiltonian lattice gauge theory~\cite{Arxiv2203_16012}, and subtracting a modified ground-state energy is a known strategy for bounding low-energy expectations~\cite{Arxiv2604_24896}. In the present context, Eq.~\eqref{eq:vacuum-cost} supplies the exact link-independent local threshold needed to establish spectral separation.

We partition the untruncated lattice Hamiltonian into an uncoupled reference operator and a local hopping perturbation,
\begin{equation}
 \begin{aligned}
  &H_a=H_R+V_l, \\
  &H_R=H_{\rm env}+\kappa_aE_l^2+mJ_a, \\
  &V_l=-J_a\sum_{p\ni l}C_p,
 \end{aligned}
 \label{eq:split}
\end{equation}
where $\|V_l\|\le mJ_a$. Here $e_{\rm env}$ is the ground energy on the full environment torus, the same energy that enters the vacuum comparison in Eq.~\eqref{eq:vacuum-cost}. On any physical fiber with fixed link flux $E_l=k$, the local and planar flux constraints force the environment degrees of freedom to provide the return circulation. Expanding the kinetic quadratic form about the least-norm real flow and applying the diamagnetic inequality to the shifted environment form yields $H_{\rm env}\ge e_{\rm env}+\kappa_a(c_N-1)k^2$ on that fiber (Appendix~\ref{app:flow}). We denote by $Q_l(j)=\mathbf1_{\{|E_l|>j\}}$ the projector onto high-flux configurations on link $l$, setting $Q_l(-1)=I$. Restricted to the physical sector, $H_R$ commutes with $Q_l(j)$, and both $H_a$ and $H_R$ share the operator domain of the kinetic energy. On the respective ranges of $Q_l(j)$ and $P_*$, we obtain
\begin{equation}
 \begin{aligned}
 A_j&=H_R|_{\operatorname{Ran}Q_l(j)}
 \ge e_{\rm env}+mJ_a+\kappa_a c_N(j+1)^2,\\
 B&=H_a|_{\operatorname{Ran}P_*}\le e_{\rm env}+mJ_a+E_* .
 \end{aligned}
 \label{eq:separated-spectra}
\end{equation}
Consequently, the positivity condition $d_j>0$ guarantees an energy gap of at least $\kappa_a d_j$ separating the reference high-flux spectrum from the low-energy window. To bound the resulting tail projection, we apply an ordered-spectrum Sylvester estimate~\cite{AlbeverioMotovilovShkalikov2009}. To keep the presentation self-contained, we record the relevant special case and its direct verification. Let $A$ and $B$ be self-adjoint operators with $B$ bounded, and suppose there exist a real scalar $b$ and a separation parameter $\delta>0$ such that $A\geq b+\delta$ and $B\leq b$. If a bounded operator $X$ maps the domain of $B$ into the domain of $A$ and satisfies the intertwining equation $AX-XB=Y$ for bounded $Y$, then
\begin{equation}
 X=\int_0^\infty e^{-s(A-b)}Y e^{s(B-b)}\,ds,
 \qquad \|X\|\leq\frac{\|Y\|}{\delta}.
 \label{eq:sylvester}
\end{equation}
This representation follows by integrating the derivative of $F(s)=e^{-s(A-b)}Xe^{s(B-b)}$ on $[0,\infty)$, noting that $F'(s)=-e^{-s(A-b)}Ye^{s(B-b)}$ and that $\|F(s)\|\leq e^{-s\delta}\|X\|\to0$ as $s\to\infty$. In our application, $\operatorname{Ran}P_*$ is contained in the domain of $H_a$, which is also the domain of $H_R$. Because $Q_l(j)$ preserves this domain, the differentiation and integral formula hold for the unbounded operator $A_j$ without decomposing into individual eigenstates or introducing rank factors.

Setting $Q_j=Q_l(j)$ and $r_j=\|Q_jP_*\|$, the operator $X=Q_jP_*$ viewed on the range of the window projector satisfies the intertwining equation $A_jX-XB=-Q_j V_lP_*$. Because every plaquette adjacent to link $l$ increments or decrements its flux by exactly one unit, the hopping interaction obeys the selection rule
\begin{equation}
 Q_j V_l=Q_j V_l Q_{j-1}.
 \label{eq:one-step}
\end{equation}
Applying Eq.~\eqref{eq:sylvester} to this single-step transition yields the reference bound $r_j\le (A_g/d_j)r_{j-1}$ whenever $d_j>0$.

Alternatively, retain the incident plaquettes within the high-flux block. The compression $A'_j=Q_jH_aQ_j$ on its physical range is self-adjoint on the domain of the projected kinetic operator, as $Q_j$ reduces the kinetic energy and the magnetic potential is bounded. Since the incident plaquette contributions $J_a(1-C_p)$ are positive semidefinite, the fiber lower bound implies $A'_j\ge e_{\rm env}+\kappa_a c_N(j+1)^2$, which separates this block from the window by at least $\kappa_a(d_j-A_g)$ when $d_j>A_g$. In the Sylvester equation $A'_jX-XB=-Q_jH_a(1-Q_j)P_*$, the driving term receives contributions exclusively from the input shell $|E_l|=j$. For any shell index $j\ge1$, the transitions $j\to j+1$ and $-j\to -j-1$ connect mutually orthogonal input and output subspaces. Each branch is a sum of four shifts with coefficients of magnitude $J_a/2$. The crossing is their direct sum, so its norm is at most $2J_a$. At $j=0$ the branches share the input subspace $E_l=0$, and this direct-sum argument does not apply. Because the input shell is a subprojector of $Q_{j-1}$, applying Eq.~\eqref{eq:sylvester} for $j\ge1$ and $d_j>A_g$ yields the compressed factor $r_j\le A_g r_{j-1}/[2(d_j-A_g)]$.

Both estimates control $r_j$ in terms of the identical predecessor norm $r_{j-1}$. Taking the smaller applicable factor at each shell, including monotonicity $r_j\le r_{j-1}$, we obtain the recurrence
\begin{equation}
 r_j\le f_j r_{j-1},\qquad r_{-1}=1 .
 \label{eq:recursion}
\end{equation}
The compressed estimate improves upon the uncoupled reference bound precisely when $d_j>2A_g$ and $j\ge1$. Unrolling this relation from $j=0$ through $\Lambda$ proves Eq.~\eqref{eq:local-tail}. Quadratic electric energies and nearest-flux hops also appear in perturbative truncation analyses~\cite{Arxiv2508_00061}. The two bounds here apply directly to the full spectral window. Without Gauss or global flux constraints the same comparisons hold with coefficient one. The factor $c_N$ proved here uses both constraints.

The operator-norm bound in Eq.~\eqref{eq:local-tail} differs in structure from standard matrix-entry localization results for spectral projectors~\cite{BenziBoitoRazouk2013} and from Green-matrix or eigenvector estimates for block Jacobi operators~\cite{JanasNabokoSilva2018,JanasNabokoSilva2020}. The cited block-Jacobi results allow infinite-dimensional fibers but assume bounded individual block entries. Resolving $E_l$ expresses the original lattice Hamiltonian as a block-tridiagonal operator indexed by electric flux. Each fiber contains the remaining links, and its diagonal block includes the unbounded environment Hamiltonian $H_{\rm env}$. Extending bounded-block Jacobi theorems to this setting requires checking operator domains and uniformity. With the domain compatibility verified above, the ordered energy separation in Eq.~\eqref{eq:separated-spectra} gives the standard Sylvester bound for $Q_l(\Lambda)P_*$ from $\operatorname{Ran}P_*$ to $\operatorname{Ran}Q_l(\Lambda)$. This controls the original spectral-window tail without a prefactor depending on the window dimension.

\begin{proposition}[Sharpness at fixed lattice size]\label{prop:sharp}
Fix the lattice geometry, the grid spacing, and the coupling $g>0$. For each link $l$, the true tail amplitude satisfies
\begin{equation}
 \log\|Q_l(\Lambda)P_*\|=-2\Lambda\log\Lambda+O(\Lambda),
 \qquad \Lambda\to\infty .
 \label{eq:true-tail}
 \end{equation}
The same logarithmic asymptotic behavior governs both the vacuum tail $\|Q_l(\Lambda)\Omega\|$ and the global hard-wall defect $\|(1-P_\Lambda)P_*\|$.
\end{proposition}
For a single Mathieu rotor, high-order Fourier coefficients and Rayleigh--Ritz eigenvalue errors already exhibit distinct factorial scales~\cite{WolfDLMFMathieu,VolkmerMathieuRitz}. Here the lower bound matches the leading logarithmic rate of the tail: it follows from the amplitudes of repeated plaquette-loop configurations in the interacting vacuum.

The angle-space vacuum is strictly positive on the configuration torus. Its Haar integral, the all-zero electric coefficient, therefore satisfies $c_0>0$. Replacing each electric-basis amplitude by its absolute value preserves the kinetic form, its domain, and support in the Gauss-law, zero-winding sector; moreover, because every plaquette hop carries a negative off-diagonal matrix element, the magnetic expectation value cannot increase. The resulting state must therefore also be a ground state. Vacuum uniqueness together with $c_0>0$ fixes the global phase, establishing that all electric-basis coefficients are nonnegative.

Now consider an elementary plaquette $p$ containing $l$ and set $c_k=\langle k\partial p|\Omega\rangle$. The closed loop $k\partial p$ is contractible, possesses zero winding, and has kinetic energy $4\kappa_a k^2$. In the ground-state eigenvalue equation evaluated at this loop state, the adjacent configuration $(k-1)\partial p$ contributes $J_a c_{k-1}/2$, while all other neighboring configurations contribute nonnegatively. Applying the bound $0\le E_0\le n_pJ_a$, which follows from positivity and the zero-flux trial state, we obtain for every $k\ge1$
\begin{equation}
 \begin{aligned}
  &c_k\geq\frac{J_a c_{k-1}}{2(4\kappa_a k^2+n_pJ_a-E_0)} \geq \frac{b}{k^2}c_{k-1}, \\
  &b=\frac{J_a}{2(4\kappa_a+n_pJ_a)}>0 .
 \end{aligned}
 \label{eq:vacuum-recursion}
\end{equation}
Iterating to $k=\Lambda+1$ gives $c_0 b^k/(k!)^2\le\|Q_l(\Lambda)\Omega\|\le\|Q_l(\Lambda)P_*\|\le R_\Lambda$. By commutativity of the link projections, the global defect is bounded below by the single-link tail and above by $\sqrt{n_l}R_\Lambda$. Applying Stirling's approximation to both sides yields Eq.~\eqref{eq:true-tail}; the subleading terms linear in $\Lambda$ need not coincide.

This factorial bound cannot be deduced merely from an upper bound on the expectation value of the Hamiltonian. To see why, consider a closed electric flux loop state $|e_M\rangle$ carrying a positive integer flux $M$ around an elementary plaquette on a fixed lattice. This loop state satisfies Gauss's law, carries zero net flux across the original electric-flux cuts, and has expectation value
\begin{equation}
 \langle e_M|H_a|e_M\rangle=4\kappa_aM^2+n_pJ_a.
 \label{eq:loop-energy}
\end{equation}
For any positive target excitation energy $E_*>0$ and sufficiently large $M$, the statistical mixture $(1-w_M)|\Omega\rangle\langle\Omega|+w_M|e_M\rangle\langle e_M|$, with $|\Omega\rangle$ the ground state and $w_M=E_*/(4\kappa_aM^2+n_pJ_a-E_0)$, has an exact mean excitation of $E_*$. For any link $l$ of the plaquette, the probability of $|E_l|>M-1$ is $(1-w_M)\|Q_l(M-1)|\Omega\rangle\|^2+w_M\geq w_M\sim M^{-2}$, ruling out factorial suppression (see Appendix~\ref{app:auxiliary}). Thus, on the same underlying lattice regulator, spectral-window support is a strictly stronger requirement than an expectation-value constraint.
\subsection{A weighted tail at weak coupling}\label{sec:weighted}

The shell product is fully explicit, yet its first contractive factor recedes to flux magnitudes of order $g^{-2}$ as $g\to0$. Fang and Tong establish a weak-coupling bound on the same window norm by analyzing an exponentially weighted state~\cite[Proposition~1]{FangTong2026}. Their three-dimensional periodic comparison focuses on the neutral sector analyzed here, while their general theorem extends beyond it. We use their local vacuum comparison and whole-window weighted estimate with a modified weight; the compensating flow raises the electric penalty entering that estimate.

Set $\tau=g^2$, $A_{\rm w}=4+aE_*$, and, for $b>0$ and $y\ge0$, define the action $\mathcal I_b(y)=\int_0^y\operatorname{arcosh}(1+s^2/(2b))\,ds$. Write $q_j^{\rm FT}$ for the explicit bound of Fang and Tong in three dimensions, and $q_{c,j}$ for the modified bound with electric cost coefficient $c$. At nonnegative integer shell index $j$ the two formulas read

\begin{equation}\begin{aligned} F(j)&=\mathcal I_4(\tau j)/(2\tau),\\ G_c(j)&=\mathcal I_8(\sqrt c\,\tau j)/(\sqrt c\,\tau),\\ q_j^{\rm FT}&=\sqrt2\exp[A_{\rm w}-F(j)],\\ q_{c,j}&=\sqrt2\exp[\sqrt{2/c}\,A_{\rm w}-G_c(j)]. \end{aligned}\label{eq:weighted-direct}\end{equation}

\begin{proposition}[Weighted electric-flux tail]\label{prop:weighted}Within the physical subspace specified in Theorem~\ref{thm:main}, the tail bound $\|Q_l(j)P_*\|\le q_{c,j}$ holds for both $c=1$ and $c=c_N$. Provided that the Fang--Tong certificate satisfies $q_j^{\rm FT}\le1$, the strict improvement $q_{c_N,j}<q_j^{\rm FT}$ is guaranteed.\end{proposition}

The rescaled Hamiltonian $\mathcal K=a\tau H_a$ has electric coefficient $\tau^2/2$ and magnetic coefficient one. In these units Fang and Tong bound the difference $a\tau(E_0-e_{\rm env})$ by $4\tau$ from above~\cite[Lemma~3]{FangTong2026}. The deleted-link operator is $a\tau H_{\rm env}$ on the environment torus of Eq.~\eqref{eq:environment}. Its ground energy is precisely the reference retained by our fiber estimate, not the energy of an independently constrained trial state. This identifies the energy origin needed to combine the estimates. Appendix~\ref{app:weight} proves the proposition, including the comparison of the prefactors.

The direct estimate can also initialize further shell steps. If $\rho_{-1}=1$, define $\rho_j=\min\{1,q_j^{\rm FT},q_{c_N,j},f_j\rho_{j-1}\}$ for $j\ge0$. Induction in Eq.~\eqref{eq:recursion} gives $\|Q_l(j)P_*\|\le\rho_j\le R_j$. The construction is never weaker than the original product certificate. The $q_j^{\rm FT}$ entry is redundant once both one and $q_{c_N,j}$ are available; propagating the published bound alone, however, need not give the same cutoff as taking a minimum only at the final shell.

%% file: sections/spectroscopy.tex
\section{Energy errors from the two sides of the hard wall}
\label{sec:spectroscopy}

A plaquette hop changes the electric-flux integer by one on each of its four links, with signs set by orientation. To cross the wall, it must join a configuration on the last retained shell to one on the first discarded shell. The energy estimate uses the amplitudes on both shells. In Rayleigh--Ritz approximation the error in the Ritz values is already known to be quadratic in the subspace angle near an invariant subspace~\cite{KnyazevArgentatiRitz}; in a tridiagonal problem, Volkmer estimates a matrix element using the boundary coupling times two eigenvector coefficients, one at the last retained index and one at the first omitted index~\cite[Lemma~3, p.~44]{VolkmerMathieuRitz}. Here a plaquette hop can cross the wall only between adjacent flux shells. Bounds on the two shell amplitudes therefore control the crossing bilinear form uniformly on the original spectral window.

\subsection{Ordered levels in the original window}
\label{sec:spectral-accuracy}

\begin{corollary}[Ordered energy shifts]
List the eigenvalues $E_k$ of $H_a$ and $E_k^\Lambda$ of the compression $H_{a,\Lambda}$ in the same physical zero-flux sector, starting at $k=0$ and counting multiplicities. Set $P=P_\Lambda$, $Q=1-P$, and $\eta=\|QP_*\|$. If $\eta<1$, the $k$th compressed eigenvalue exists for each $k$ with $E_k\le E_0+E_*$, and its shift satisfies
\begin{equation}
 0\le E_k^\Lambda-E_k
 \le \frac{\|P_*P H_a QP_*\|}{1-\eta^2}.
 \label{eq:spectral-accuracy}
\end{equation}
For $n_lR_\Lambda^2<1$ an explicit upper bound on this shift is
\begin{equation}
 \begin{aligned}
  &\varepsilon_\Lambda=\frac{2J_an_p R_\Lambda s_\Lambda}{1-n_lR_\Lambda^2}, \\
  &s_0=1,\quad s_\Lambda=\min\{1,2R_{\Lambda-1}\}\quad(\Lambda\ge1).
 \end{aligned}
 \label{eq:spectral-accuracy-explicit}
\end{equation}
The corresponding excitation energies obey $|(E_k^\Lambda-E_0^\Lambda)-(E_k-E_0)|\le\varepsilon_\Lambda$.
\end{corollary}

\begin{proof}
For a fixed plaquette $p$, let $P_p(j)$ project onto fluxes of magnitude at most $j$ on its four links. Write $P=P_p(\Lambda)P_{\rm out}$, where $P_{\rm out}$ restricts all other links. It commutes with the plaquette hopping operator $C_p$. For $\Lambda\ge1$, a hop from outside $P_p(\Lambda)$ to inside it must end at flux $\pm\Lambda$ on at least one perimeter link. With $S_p=P_p(\Lambda)-P_p(\Lambda-1)$ this gives
\begin{equation}
 PC_p Q=S_p P_{\rm out} C_p [1-P_p(\Lambda)]P_{\rm out}.
 \label{eq:spectral-shells}
\end{equation}
For any unit window vectors $\phi,\psi$, the inner and outer factors satisfy
$\|S_pP_{\rm out}\phi\|\le s_\Lambda$ and
$\|[1-P_p(\Lambda)]P_{\rm out}\psi\|\le 2R_\Lambda$.
Indeed, $1-P_p(j)\le\sum_{l\in\partial p}Q_l(j)$; apply Eq.~\eqref{eq:local-tail} to the original window vectors after dropping the commuting projection $P_{\rm out}$ from each norm. At $\Lambda=0$ use $P_p(0)$ on the inner side and its norm bound one. The kinetic term and the constant part of the plaquette energy are diagonal, so only the hopping terms couple $P$ and $Q$. Summing their bilinear bounds and applying the same projection inequality to all links gives
\[
 \|P_*P H_a QP_*\|\le 2J_an_pR_\Lambda s_\Lambda,\qquad
 \eta^2\le n_l R_\Lambda^2 .
\]

For each such $k$, let $U$ have the first $k+1$ orthonormal eigenvectors of $H_a$ as columns, and let $B=\operatorname{diag}(E_0,\ldots,E_k)$. Thus $H_aU=UB$. The projected columns $F=PU$ have Gram matrix $G=F^\dagger F\succeq(1-\eta^2)I$. They span an admissible trial space of dimension $k+1$: $P$ preserves the sector and the kinetic-energy domain. From $H_a U=UB$, the energy matrix on these columns is
\begin{equation}
 F^\dagger H_aF=GB-C,\qquad C=U^\dagger P H_a Q U .
 \label{eq:spectral-variational-identity}
\end{equation}
Let $\mu$ be the largest Ritz value in this trial space, and choose its coefficient vector $v$ to have Euclidean norm one. Since $F^\dagger H_a F$ is Hermitian and $G>0$, $\mu$ is real. The generalized eigenvalue equation gives $\mu v=Bv-G^{-1}Cv$. Taking the real part of the scalar product with $v$ bounds $\mu$ by $E_k+\|C\|/(1-\eta^2)$, without requiring $G^{-1}C$ to be Hermitian. By min--max, $E_k^\Lambda\le\mu$, while restricting the original variational problem to the wall yields $E_k^\Lambda\ge E_k$. Since $\|C\|\le\|P_*P H_a QP_*\|$, the two energy bounds follow. Subtracting the ground-state shift from the $k$th shift takes the difference of two numbers in $[0,\varepsilon_\Lambda]$, which proves the excitation-energy bound.
\end{proof}

\begin{corollary}[Window membership from the finite spectrum]
\label{cor:finite-spectrum}
In the same sector and with the same energy origin, let $\lambda_k=E_k^\Lambda$ be ordered with multiplicity and let $\omega_k=\lambda_k-\lambda_0$. Write $\epsilon=\varepsilon_\Lambda(E_*)$ and $\epsilon_{\rm vac}=\varepsilon_\Lambda(0)$ for the bounds in Eq.~\eqref{eq:spectral-accuracy-explicit}, evaluated with positive denominators. If $\omega_k\le E_*-\epsilon_{\rm vac}$, the original level lies in the window and satisfies
\[
\begin{aligned}
 \lambda_k-\epsilon &\le E_k\le\lambda_k,\\
 \max(0,\omega_k-\epsilon)&\le E_k-E_0\le\omega_k+\epsilon_{\rm vac}.
\end{aligned}
\]
Let $\mathcal N_\Lambda(x)$ count the finite-wall excitation energies at most $x$, including multiplicities, with count zero for $x<0$. Then
\[
 \mathcal N_\Lambda(E_*-\epsilon_{\rm vac})
 \le {\rm rank}\,P_*
 \le \mathcal N_\Lambda(E_*+\epsilon).
\]
\end{corollary}

\begin{proof}
Min--max gives $E_k\le\lambda_k$. Independently, the vacuum enclosure yields $E_0\ge\lambda_0-\epsilon_{\rm vac}$. Combining them proves membership when $E_k-E_0\le\omega_k+\epsilon_{\rm vac}\le E_*$; we can now apply the window bound to enclose $E_k$. Subtracting the vacuum interval gives the excitation bounds. Thus, every finite-wall index meeting the threshold contributes to the lower count. Conversely, positivity of the Gram denominator ensures that every original window level has a finite-wall counterpart satisfying $\lambda_k-\lambda_0\le E_k-E_0+\epsilon\le E_*+\epsilon$, which establishes the upper count.
\end{proof}

Equal finite counts determine the dimension of the original window; no isolation gap at the window edge is assumed. For numerical use, individual energy intervals require certified enclosures for the ordered finite-wall energies, and a dimension certificate requires rigorous counts at both thresholds. Neither conclusion determines the spectral support of the truncated eigenvectors.

The bounds on the two contiguous shells in Eq.~\eqref{eq:spectral-accuracy-explicit} can be evaluated either via Proposition~\ref{prop:weighted} or through the iterated bound $\rho$. Under the recursive choice, the outer factor is replaced by $\rho_\Lambda$ and the inner factor by $\min(1,2\rho_{\Lambda-1})$ (taking value $1$ when $\Lambda=0$), provided $1-n_l\rho_\Lambda^2>0$. The spectral window membership and level-counting assertions above apply these estimates at the two designated energy widths.

\subsection{Sharp rate for the absolute ground-state energy}
\label{sec:vacuum-energy-rate}

\begin{proposition}[Vacuum energy at fixed lattice size]
Fix the finite lattice, its spacing, and $g>0$. Let $\Delta_{\rm vac}=E_1-E_0$ be the vacuum gap in the physical zero-flux sector. Then
\begin{equation}
 \log\!\left(\frac{E_0^\Lambda-E_0}{\Delta_{\rm vac}}\right)
 =-4\Lambda\log\Lambda+O(\Lambda),\;\Lambda\to\infty .
 \label{eq:vacuum-energy-rate}
\end{equation}
\end{proposition}

\begin{proof}
The positivity-improving heat kernel in the angle representation gives a unique ground state in this sector (Sec.~\ref{sec:tail}). Compact resolvent at fixed lattice size then gives $\Delta_{\rm vac}>0$. Every unit wall state $w$ in this sector satisfies $|\langle\Omega,w\rangle|\le\|P\Omega\|$. The spectral theorem gives $\langle w,H_aw\rangle-E_0\ge\Delta_{\rm vac}(1-|\langle\Omega,w\rangle|^2)$. Minimizing over $w$ yields
\begin{equation}
 E_0^\Lambda-E_0\ge\Delta_{\rm vac}\|(1-P_\Lambda)\Omega\|^2 .
 \label{eq:vacuum-energy-lower}
\end{equation}
For any link $l$, $\|(1-P_\Lambda)\Omega\|\ge\|Q_l(\Lambda)\Omega\|$. The latter norm is nonzero at every finite cutoff by Eq.~\eqref{eq:vacuum-recursion}. Its logarithm is $-2\Lambda\log\Lambda+O(\Lambda)$ by Proposition~\ref{prop:sharp}; squaring it in Eq.~\eqref{eq:vacuum-energy-lower} gives the lower logarithmic coefficient $-4$. For $\Lambda\ge1$ with $d_\Lambda>2A_g$, in particular for all sufficiently large $\Lambda$, Eq.~\eqref{eq:tail-product} gives $R_{\Lambda-1}/R_\Lambda=2(d_\Lambda-A_g)/A_g$. This quadratic factor does not change the leading logarithm of the squared tail in Eq.~\eqref{eq:spectral-accuracy-explicit}, furnishing the upper side of Eq.~\eqref{eq:vacuum-energy-rate}.
\end{proof}

The lower bound uses the vacuum gap at this fixed volume, not a separation at the upper edge of the low-energy window. It makes no claim that the error of each excited level or of an excitation energy has a matching lower bound. The two small shell amplitudes in Eq.~\eqref{eq:spectral-shells} enter between window vectors on both sides; in the equation of motion that follows, only the input lies in the window.

%% file: sections/dynamics.tex
\section{Dynamics, bounded readout, and preparation leakage}
\label{sec:dynamics}

Because the untruncated Hamiltonian $H_a$ commutes with the spectral window projector $P_*$, the low-energy subspace is strictly invariant under the exact time evolution operator $U(t)=e^{-itH_a}$. This invariance keeps the static tail bound valid along the original trajectory, without requiring the truncated evolution to preserve the window. Integrating the forcing then gives the finite-time comparison below. Setting $P=P_\Lambda$ and $Q=1-P$, the projected state $PU(t)\psi$ for any $\psi\in\operatorname{Ran}P_*$ satisfies the differential equation
\begin{equation}
 i\partial_t(PU(t)\psi)=H_{a,\Lambda}PU(t)\psi+PH_aQU(t)\psi.
 \label{eq:forcing}
\end{equation}
The electric kinetic operator is completely diagonal in the charge basis, so $P\sum_l E_l^2 Q=0$. Consequently, the off-diagonal driving term arises solely from the magnetic plaquette sum. For each plaquette $p$, we factor the global cutoff into $P=P_p P_{\rm out}$, where $P_l(\Lambda)=\mathbf1_{[-\Lambda,\Lambda]}(E_l)$ and $P_p=\prod_{l\in p}P_l(\Lambda)$ acts on the four perimeter links and $P_{\rm out}=\prod_{l\notin p}P_l(\Lambda)$ acts on the remaining links. Because $C_p$ does not alter links outside $p$, it commutes with $P_{\rm out}$, yielding the identity
\begin{equation}
 \begin{aligned}
  &PC_pQ=P_pC_p(1-P_p)P_{\rm out}, \\
  &\|PC_pQP_*\|\leq2R_\Lambda.
 \end{aligned}
 \label{eq:plaquette-forcing}
\end{equation}
The operator $P_{\rm out}$ is an orthogonal projection and therefore contractive. Although $P_{\rm out}$ does not generally commute with $P_*$, it commutes with the local projection $1-P_p$. Combined with the operator inequality $1-P_p\leq\sum_{l\in p}Q_l(\Lambda)$ and the norm bound $\|C_p\|\leq1$, this property bounds the plaquette driving norm by $2R_\Lambda$. Summing this contribution over all $n_p$ plaquettes gives the bound on the total forcing, while the initial truncation defect is bounded by the square root of the link volume:
\begin{equation}
 \begin{aligned}
  &\|PH_aQP_*\|\leq2J_an_pR_\Lambda, \\
  &\|(1-P_\Lambda)P_*\|\leq\sqrt{n_l}R_\Lambda.
 \end{aligned}
 \label{eq:forcing-bounds}
\end{equation}
Crucially, the driving term $PH_aQ$ acts on the exterior component $Q\psi$, so the relevant electric cutoff index is $\Lambda$. A two-way commutator estimate based on $[H_a,P_\Lambda]$ would probe both sides of the truncation boundary and introduce the weaker tail index $\Lambda-1$.

Integrating Eq.~\eqref{eq:forcing} via variation of constants and applying the bounds in Eq.~\eqref{eq:forcing-bounds} gives the discrepancy for the unnormalized projected state,
\begin{equation}
 \|(U(t)-e^{-itH_{a,\Lambda}}P_\Lambda)P_*\|
 \leq(\sqrt{n_l}+2|t|J_an_p)R_\Lambda.
 \label{eq:unnormalized-dynamics}
\end{equation}
Now consider a normalized vector $\psi\in\operatorname{Ran}P_*$. When the projection defect $q_\psi=\|(1-P_\Lambda)\psi\|$ is strictly less than one, the normalized state on the truncated space is $\psi_\Lambda=P_\Lambda\psi/\|P_\Lambda\psi\|$. The norm distance between the unnormalized projection and its normalized counterpart is bounded by $1-\sqrt{1-q_\psi^2}\leq q_\psi$. Combining this geometric bound with Eq.~\eqref{eq:unnormalized-dynamics} yields the uniform trajectory bound in Eq.~\eqref{eq:trajectory}. This derivation applies to the entire invariant subspace simultaneously, avoiding any need to propagate or sum individual energy eigenstates.

For any bounded observable $O$ with $\|O\|\leq1$, we associate the compressed operator $O_\Lambda=P_\Lambda OP_\Lambda$ acting on $\operatorname{Ran}P_\Lambda$. Standard algebraic rearrangement of expectation values yields
\begin{equation}
 \begin{aligned}
  &\sup_{|t|\leq T} \bigl|\langle U(t)\psi,O U(t)\psi\rangle \\
  &\quad{}-\langle e^{-itH_{a,\Lambda}}\psi_\Lambda, O_\Lambda e^{-itH_{a,\Lambda}}\psi_\Lambda\rangle\bigr| \\
  &\leq2B_NR_\Lambda.
 \end{aligned}
 \label{eq:observable-error}
\end{equation}
A concrete family of bounded, gauge-invariant observables is given by the Weyl operators associated with transverse spatial smearing functions. Let $f$ be the transverse test profile defined by $f_1(x)=\alpha_f\sin(2\pi x_2/\ell)$ and $f_2=f_3=0$, with constant amplitude $\alpha_f\in\mathbb R$ restricted to $0<g\ell|\alpha_f|<3\pi$. The corresponding unitary operator is
\begin{equation}
 W_f(a)=\exp\!\left[iga\sum_{x,i}f_i(ax)E_{x,i}\right].
 \label{eq:weyl-readout}
\end{equation}
This operator commutes with the Gauss law generators, the global flux constraints, and the hard-wall cutoff $P_\Lambda$. Its real and imaginary parts define bounded self-adjoint observables. On the full physical zero-flux Hilbert space, $W_f(a)$ does not reduce to the identity operator (Appendix~\ref{app:auxiliary}). 

With the normalization $E_{{\rm phys},l}=gE_l/a^2$, the lattice exponent $ga\sum_{x,i}f_i(ax)E_{x,i}$ in Eq.~\eqref{eq:weyl-readout} directly discretizes the continuum spatial smear $\int f\cdot E_{\rm phys}\,d^3x$. The unitary operator $W_f(a)$ therefore serves as the gauge-invariant lattice Weyl probe realizing this transverse field profile.

Finally, an initial state prepared in a simulation may only approximate the target low-energy window. Let $\psi$ be a normalized state in the same zero-electric-flux physical sector whose high-energy leakage amplitude is quantified by
\begin{equation}
 \delta=\|(1-P_*)\psi\|.
 \label{eq:leakage}
\end{equation}
Decomposing the state as $\psi=P_*\psi+(1-P_*)\psi$, the high-energy component $(1-P_*)\psi$ contributes at most $2\delta$ to the unnormalized evolution discrepancy because $\|U(t)-e^{-itH_{a,\Lambda}}P_\Lambda\|\leq2$. The missing norm after electric-flux truncation satisfies $\|(1-P_\Lambda)\psi\|\leq\sqrt{n_l}R_\Lambda+\delta$. Whenever this quantity is strictly less than one, normalizing the truncated state produces the composite error bounds
\begin{equation}
 \begin{aligned}
  &\sup_{|t|\leq T}\|U(t)\psi-e^{-itH_{a,\Lambda}}\psi_\Lambda\| \\
  &\leq B_NR_\Lambda+3\delta, \\
  &|\Delta\langle O\rangle|\leq2B_NR_\Lambda+6\delta.
 \end{aligned}
 \label{eq:leakage-budget}
\end{equation}
Equation~\eqref{eq:leakage-budget} budgets the truncation error and state-preparation leakage as separate additive terms. Because $\delta$ measures state-vector amplitude, an initial leakage probability $p$ enters the observable expectation discrepancy via the square-root scaling $6\sqrt p$. Adjusting the electric cutoff $\Lambda$ controls the dynamical truncation defect, whereas the initial leakage $\delta$ is certified independently by the state-generation protocol.

\subsection{A linear encoding of the spectral window}

Let $\mathcal S=\operatorname{Ran}P_*$ and suppose $\eta=\|(1-P_\Lambda)P_*\|<1$. The restriction $F_\Lambda=P_\Lambda|_{\mathcal S}$ has no kernel. Its polar factor defines an isometry~\cite{Watrous2018Theory},
\begin{equation}
 \begin{aligned}
  &F_\Lambda^\dagger F_\Lambda\geq(1-\eta^2)I_{\mathcal S}, \\
  &V_\Lambda=F_\Lambda(F_\Lambda^\dagger F_\Lambda)^{-1/2},\qquad V_\Lambda^\dagger V_\Lambda=I_{\mathcal S}.
 \end{aligned}
 \label{eq:polar-encoding}
\end{equation}
Because $P_\Lambda$ preserves the Gauss and global-flux constraints, the image remains in the same physical neutral sector. In contrast to the map $\psi\mapsto P_\Lambda\psi/\|P_\Lambda\psi\|$, $V_\Lambda$ is linear on the entire window. These maps generally give different truncated states.

The singular values of $F_\Lambda$ belong to $[\sqrt{1-\eta^2},1]$, and hence
\begin{equation}
 \|V_\Lambda-F_\Lambda\|=1-\sqrt{1-\eta^2}\leq\eta.
 \label{eq:polar-distance}
\end{equation}
Combining this inequality with Eq.~\eqref{eq:unnormalized-dynamics} gives, with the truncated output naturally included in the original Hilbert space,
\begin{equation}
 \sup_{|t|\leq T}\|U(t)|_{\mathcal S}-e^{-itH_{a,\Lambda}}V_\Lambda\|
 \leq B_NR_\Lambda.
 \label{eq:encoded-dynamics}
\end{equation}
A single linear encoding thus governs the dynamics of the entire low-energy window with the uniform error bound $B_N R_\Lambda$.

For density operators $\rho$ supported on $\mathcal S$, let $\Phi_t(\rho)=U(t)\rho U(t)^\dagger$ and $\Psi_t(\rho)=e^{-itH_{a,\Lambda}}V_\Lambda\rho V_\Lambda^\dagger e^{itH_{a,\Lambda}}$, with both channels mapping into the original Hilbert space. Their diamond-norm difference satisfies
\begin{equation}
 \sup_{|t|\leq T}\|\Phi_t-\Psi_t\|_\diamond\leq 2B_NR_\Lambda.
 \label{eq:channel-distance}
\end{equation}
For any auxiliary reference and trace-class input $X$ on $\mathcal S\otimes\mathcal H_{\rm aux}$, set $A=(U(t)|_{\mathcal S})\otimes I_{\rm aux}$ and $B=(e^{-itH_{a,\Lambda}}V_\Lambda)\otimes I_{\rm aux}$, using the common output inclusion above. Then $AXA^\dagger-BXB^\dagger=(A-B)XA^\dagger+BX(A^\dagger-B^\dagger)$. The trace-ideal inequality bounds each term by $\|A-B\|\|X\|_1$, proving Eq.~\eqref{eq:channel-distance} after taking the supremum over inputs and references. This proof applies to the infinite-dimensional output; Ref.~\cite{Watrous2018Theory} reviews the standard finite-dimensional channel-norm interpretation. Through the channel diamond norm, this dynamical guarantee extends directly to mixed states and arbitrary reference-entangled inputs supported on $\mathcal S$.

\subsection{Inputs prepared inside the hard wall}
Let $J$ be an isometry from a finite-dimensional logical space into the physical Hilbert space satisfying $P_\Lambda J=J$, with leakage parameter $\delta_J=\|(1-P_*)J\|$. The exact and truncated dynamics can both be initiated directly from $J$. Setting $C_N(T)=\sqrt{n_l}+2TJ_an_p$ and viewing the truncated output within the full Hilbert space, decomposing $(U(t)-e^{-itH_{a,\Lambda}}P_\Lambda)J$ across $P_*$ and applying Eq.~\eqref{eq:unnormalized-dynamics} bounds the operator-norm difference of the two isometries by $C_N(T)R_\Lambda+2\delta_J$ for $|t|\le T$. The high-energy component costs at most $2\delta_J$, since both $U(t)$ and $e^{-itH_{a,\Lambda}}P_\Lambda$ are contractions; the latter projection must still act on $(1-P_*)J$. The trace-ideal argument from Eq.~\eqref{eq:channel-distance} then bounds the corresponding unhalved channel diamond distance by
\begin{equation}
 2C_N(T)R_\Lambda+4\delta_J .
 \label{eq:trial-channel}
\end{equation}
For any normalized physical vector $\psi$ prepared inside the wall, the same bound holds for the readout error of any $\|O\|\le1$, with $\delta_J$ replaced by $\|(1-P_*)\psi\|$. No input renormalization is needed.

For $\operatorname{Ran}J\subset D(H_a)$, leakage can also be bounded from the trial residual. Set $A=J^\dagger H_aJ$ and $\mathcal R=H_aJ-JA$, in the physical zero-flux sector. If $A\le bI$ and the complementary spectrum is bounded below by $d>b$, the ordered Sylvester equation for $X=(1-P_*)J$ has right-hand side $(1-P_*)\mathcal R$. Equation~\eqref{eq:sylvester} gives $\delta_J\le\min\{1,\|\mathcal R\|/(d-b)\}$. For finite flux support, evaluate the residual on the trial support and on all configurations reached from it by one plaquette hop, including those outside the preparation wall.

One possible lower edge uses an auxiliary wall $M$, not necessarily the preparation wall $\Lambda$. Given a certified lower bound $L_M\le E_0^M$, evaluate Eq.~\eqref{eq:spectral-accuracy-explicit} at width zero and wall $M$, with positive Gram denominator, to obtain $\varepsilon_{{\rm vac},M}$. Then $\Delta_M=L_M-\varepsilon_{{\rm vac},M}+E_*\le E_0+E_*$ can serve as $d$ if $\Delta_M>b$. Both wall calculations use the same sector and the energy convention of Eq.~\eqref{eq:hamiltonian}. A variational upper estimate on $E_0^M$ cannot replace $L_M$. At $E_*=0$ we have $\Delta_M\le E_0\le b$, so this particular lower edge cannot certify separation. The example below uses a stronger complementary edge obtained directly from its band structure.

%% file: sections/coherent-example.tex
\section{Coherent plaquette dynamics on the cubic torus}
\label{sec:coherent-example}

We construct a nonvacuum spectral window with a dynamical plaquette readout on the original three-dimensional periodic cubic lattice, using electric-loop states of the standard gauge-invariant representation~\cite{KogutSusskind1975,KaplanStryker2020}.

Fix $N=3$, $\ell=1$, $g=6$, and a spectral-window width $E_*=270$ above the exact vacuum energy. There are $81$ links and $81$ plaquettes, with $\kappa_a=54$ and $J_a=1/12$. The dimensionless choices $n_pJ_a/\kappa_a=1/8$ and $E_*/\kappa_a=5$ keep the magnetic perturbation budget small relative to the kinetic spacing and place the window width between the kinetic scales $4\kappa_a$ and $6\kappa_a$. The electric kinetic operator in the neutral zero-flux sector has a unique zero-electric state, followed by $162$ oriented unit-plaquette states at energy $216$; its next level is at least $324$. At this perturbation strength, the exact vacuum and the first excited band are spectrally separated from each other and from higher excitations. Appendix~\ref{app:coherent-proof} proves that $P_*$ contains precisely this vacuum and the whole $162$-dimensional band, including any internal degeneracies.

\subsection{An interacting-window wavepacket}
Let $P_{\rm band}=P_*-|\Omega\rangle\langle\Omega|$ project onto that band. For any fixed elementary plaquette, the self-adjoint plaquette observable $C_p$ has a certified transition strength
\begin{equation}
 \begin{aligned}
  &z=\|P_{\rm band}C_p\Omega\|\geq\frac{4\sqrt7}{15}-\frac1{279}>0.701, \\
  &\chi=\frac{P_{\rm band}C_p\Omega}{z},\qquad \psi=\frac{\Omega+\chi}{\sqrt2}.
 \end{aligned}
 \label{eq:coherent-wavepacket}
\end{equation}
The state $\psi$ coherently combines the vacuum with an excited wavepacket in the original window. Its plaquette expectation $M(t)=\langle U(t)\psi,C_pU(t)\psi\rangle$ changes by an order-one amount:
\begin{equation}
 M(0)-M(t_0)>1.30,\qquad t_0=\frac{\pi}{216}.
 \label{eq:coherent-signal}
\end{equation}
The reference time $t_0=\pi/(4\kappa_a)$ is the half-period of the unperturbed vacuum--plaquette transition. Appendix~\ref{app:coherent-proof} controls the shift and spread of the exact frequencies, together with the nonzero transition inside the selected window, to establish the stated change between $0$ and $t_0$.

For this window $A_g=1/162$, $s_a=5$, $c_3=81/52$ and $j_0=1$. Here $d_1=4c_3-s_a=16/13>2A_g$, so the compressed factor applies already at $|E_l|=2$. At $T=1/50>t_0$,
\begin{equation}
 \begin{aligned}
 B_3&=\frac{1827}{100},\quad R_3=\frac{2197}{16426765489060},\\
 2B_3 R_3&<4.89\times10^{-9}.
 \end{aligned}
 \label{eq:coherent-budget}
\end{equation}
At $T=1/50$ the least cutoff certified for readout tolerance $10^{-3}$ drops from three to two when the return-flow cost is included, even using just the reference step. For the joint dynamical and spectral example we keep $\Lambda=3$, or seven flux levels per link. For the interacting wavepacket $\psi$ of Eq.~\eqref{eq:coherent-wavepacket}, set $\psi_\Lambda(t)=e^{-itH_{a,\Lambda}}P_\Lambda\psi/\|P_\Lambda\psi\|$. Its hard-wall plaquette readout is $M_\Lambda(t)=\langle\psi_\Lambda(t),P_\Lambda C_pP_\Lambda\psi_\Lambda(t)\rangle$. Applying Eq.~\eqref{eq:observable-error} at the two endpoints gives $M_3(0)-M_3(t_0)>1.29$.

For the same wall, $R_2=169/195813154$. The two shell bounds in Eq.~\eqref{eq:spectral-accuracy-explicit} give
\begin{equation}
 \begin{aligned}
 0\le E_k^3-E_k &< 3.12\times10^{-15},\qquad 0\le k\le162 .
 \end{aligned}
 \label{eq:coherent-spectral-accuracy}
\end{equation}
The band count comes from the untruncated spectrum. The bound holds for all $163$ ordered levels, and in absolute value for their excitation-energy errors. Taking $E_*=0$ instead gives a vacuum-only upper bound of $6.27\times10^{-22}$. The vacuum lower bound at this wall is about $9.45\times10^{-34}$ (Appendix~\ref{app:coherent-proof}); matching logarithmic rates do not make the finite-cutoff bracket tight. For arbitrary inputs in the full window the polar encoding at $T=1$ has channel error below $8.43\times10^{-9}$ by Eq.~\eqref{eq:channel-distance}.

\subsection{An input prepared from three electric configurations}
While the preceding wavepacket uses the interacting vacuum and the projection onto the excited band, an alternative input state can be constructed without either ingredient. For an oriented plaquette $p$, set $|p_\pm\rangle=U_p^{\pm1}|0\rangle$, $\chi_0=(|p_+\rangle+|p_-\rangle)/\sqrt2$, and $\psi_{\rm prep}=(|0\rangle+\chi_0)/\sqrt2$. All three configurations have zero divergence and zero winding. The bare configuration $|0\rangle$, unlike the interacting vacuum $\Omega$, has exactly zero electric flux on every link. In either loop configuration the flux has magnitude one on each of the four links of $\partial p$ and vanishes elsewhere.

Defining $D_p=\sum_{l\in\partial p}E_l^2$ and restricting $C_p$ to $|E_l|\le1$ on these four links yields the compressed operator $\widetilde C_p$. In the ordered basis $(|0\rangle,\chi_0)$, these operators act as $\operatorname{diag}(0,4)$ and $\sigma_x/\sqrt2$, respectively. Applying two gauge-invariant, four-link unitary pulses,
\begin{equation}
 e^{i\pi D_p/8}e^{-i\pi\widetilde C_p/(2\sqrt2)}|0\rangle=\psi_{\rm prep}
 \label{eq:prepare}
\end{equation}
generates this state exactly within any cutoff boundary $\Lambda\ge1$. Taking these two basis states as the columns of an isometry $J$ defines a trial qubit. The two ideal pulses prepare the prescribed vector, not an arbitrary input qubit entangled with a reference.

As established in Appendix~\ref{app:trial}, the leakage amplitude is bounded by $0.004898$ for $\psi_{\rm prep}$ and remains bounded by $0.005222$ uniformly across the trial qubit subspace. These certificates depend on the lower spectral edge of the complementary subspace, whereas executing the preparation circuit in Eq.~\eqref{eq:prepare} requires no spectral knowledge. For $\psi_{\rm prep}$ at $\Lambda=3$, the state-specific form of Eq.~\eqref{eq:trial-channel} bounds the plaquette readout error by $0.019589$ for $|t|\le T=1/50$. Since the initial expectation is unchanged, the truncated contrast from $0$ to $t_0$ exceeds $1.379$. Because the unperturbed trajectory for this particular state remains confined within the three initial configurations, direct comparison with unperturbed evolution yields a sharper contrast bound of $1.3987$, valid both in the full untruncated lattice model and for every $\Lambda\ge1$. Both contrast estimates apply specifically to $\psi_{\rm prep}$, rather than to the interacting wavepacket in Eq.~\eqref{eq:coherent-wavepacket}.

\begin{table}[tb]
\caption{Upper bounds on the unhalved diamond distance for the trial qubit $J$ at $\Lambda=3$. The second column lists the truncation term. The third column adds four times the unrounded upper bound on $\delta_J$ from Appendix~\ref{app:trial}, while the fourth exploits the specific flux support of $J$. Displayed bounds are rounded upward.}
\label{tab:trial}
\centering\small\setlength{\tabcolsep}{3pt}
\begin{tabular}{@{}rccc@{}}\toprule
$T$ & $2C_3(T)R_3$ & \shortstack{Window plus\\leakage} & Trial-specific\\
\midrule
$1/50$ & $2.480\times10^{-9}$ & $0.020886554$ & $2.55\times10^{-8}$\\
$1$    & $6.019\times10^{-9}$ & $0.020886557$ & $0.00095$\\
$10$   & $3.852\times10^{-8}$ & $0.020886590$ & $0.0095$\\
\bottomrule
\end{tabular}

\end{table}
Table~\ref{tab:trial} compares columns three and four on the same trial qubit in the same unhalved diamond norm. For all three durations, incorporating specific structural properties of $J$ yields tighter certificates. In contrast, the original window bound provides broader coverage by controlling arbitrary inputs supported within the 163-dimensional window. The time horizons in the table delimit the channel comparison; the contrast is evaluated at $t_0$.

%% file: sections/electric-square.tex
\section{Local electric-square readout}
\label{sec:electric-square}

Consider an exact-window initial state and either its normalized-projection or polar-encoded evolution. Write the corresponding states as $\psi_t$ and $\phi_t$. Both comparisons obey $\|\psi_t-\phi_t\|\leq B_NR_\Lambda$, and $\phi_t\in\operatorname{Ran}P_\Lambda$. The local clipped square $M_{l,\Lambda}=E_l^2\mathbf1_{\{|E_l|\leq\Lambda\}}$ has norm $\Lambda^2$. Its expectation difference is therefore at most $2\Lambda^2B_NR_\Lambda$. The remaining exact-state contribution is controlled by the integer-tail identity
\begin{equation}
 \begin{aligned}
  &k^2\mathbf1_{\{k>\Lambda\}} = (\Lambda+1)^2\mathbf1_{\{k>\Lambda\}} \\
  &\quad{}+\sum_{j=\Lambda+1}^{\infty}(2j+1)\mathbf1_{\{k>j\}}, \\
  &k\in\mathbb Z_{\geq0}.
 \end{aligned}
 \label{eq:square-tail-identity}
\end{equation}
Applying it to $k=|E_l|$ and using spectral invariance gives
\begin{equation}
 \begin{aligned}
  &\sup_{|t|\leq T}|\langle E_l^2\rangle_{\psi_t}-\langle E_l^2\rangle_{\phi_t}| \\
  &\leq 2\Lambda^2B_NR_\Lambda+(\Lambda+1)^2R_\Lambda^2 \\
  &\quad{}+\sum_{j=\Lambda+1}^{\infty}(2j+1)R_j^2.
 \end{aligned}
 \label{eq:electric-square-bound}
\end{equation}
These expectations are defined because the exact window lies in the kinetic-energy domain and the truncated state has finite flux support. Monotone convergence justifies the positive tail sum, whose convergence follows from $R_{j+1}/R_j=O(j^{-2})$.

At fixed duration $T>0$ and the other physical parameters, setting $\beta>2$ in the cutoff scaling of Sec.~\ref{sec:setup} drives the expectation error in Eq.~\eqref{eq:electric-square-bound} to zero as $N\to\infty$. Multiplying by $g^2/a^4$ yields the corresponding bound for the bare lattice field square $E_{{\rm phys},l}^2$, whose leading contribution scales as $N^{8-2\beta+o(1)}$, so $\beta>4$ is sufficient. Because the field operator is unbounded, extending this readout bound to state preparations with out-of-window leakage requires an explicit weighted moment condition, as norm distance alone cannot constrain arbitrarily large high-flux configurations.

%% file: sections/resources.tex
\section{Finite resources and regulator comparisons}
\label{sec:resources}

The product bound can be evaluated for a chosen error tolerance or inverted asymptotically. These two uses distinguish an explicit finite-lattice certificate from its scaling under refinement. Comparisons with other regulators then require identifying which operators and errors are actually being matched.

\subsection{Cutoff and preparation budgets}
\label{subsec:cutoff-budget}

For bounded readout with $\|O\|\leq1$ and target tolerance $\epsilon\in(0,1)$, define
\begin{equation}
 \begin{aligned}
  &\overline B_N=2\lceil\sqrt{3N^3}\rceil+\frac{6TN^4}{g^2\ell}, \\
  &\Lambda_O=\min\{\Lambda\in\mathbb Z_{\geq0}:2\overline B_NR_\Lambda\leq\epsilon\}.
 \end{aligned}
 \label{eq:finite-cutoff}
\end{equation}
Rounding the square root upward makes the criterion rational whenever the physical inputs are rational. Since $\overline B_N\geq2\sqrt{n_l}$, satisfying it also makes the normalized projected initial state well defined. $\Lambda_O$ is the least integer cutoff certified by Eq.~\eqref{eq:finite-cutoff}.

Table~\ref{tab:cutoffs} compares the energy and readout tests for the same window on each lattice. At all four sizes the energy test passes at a smaller cutoff than the readout test, although both cutoffs fit in six qubits per link in an independent-link binary encoding. The tabulated readout bounds assume an input supported in the original window. Section~\ref{sec:coherent-example} certifies excited levels in a different window at strong coupling.
\begin{table}[tb]
 \caption{Least cutoffs for ordered absolute energies ($\Lambda_E$) and bounded readout ($\Lambda_O$) certified by the shell product, for $\ell=T=E_*=1$ and $g=1/2$. The energy tolerance is $10^{-3}/\ell$ with positive Gram denominator; the readout tolerance is $10^{-3}$ with zero preparation leakage. The readout column uses the rational criterion in Eq.~\eqref{eq:finite-cutoff}; both error bounds are rounded upward.}
 \label{tab:cutoffs}
 \centering\small\setlength{\tabcolsep}{3pt}
 \begin{tabular}{@{}rcccc@{}}\toprule
 & \multicolumn{2}{c}{Ordered energies} & \multicolumn{2}{c}{Bounded readout}\\
 $N$ & $\Lambda_E$ & $\ell\varepsilon_{\Lambda_E}$ & $\Lambda_O$ & Readout bound\\
 \midrule
 8 & 19 & 0.000026126 & 23 & 0.000089386\\
 32 & 20 & 0.000107255 & 25 & 0.000130302\\
 128 & 21 & 0.000346741 & 27 & 0.000133749\\
 512 & 22 & 0.000909490 & 29 & 0.000100985\\
 \bottomrule
 \end{tabular}
\end{table}

A separately certified preparation error can be included without changing the cutoff calculation. For example, allocating half the total tolerance to each term in Eq.~\eqref{eq:leakage-budget} means
\begin{equation}
 \delta\leq\frac\epsilon{12},\quad
 p=\delta^2\leq\frac{\epsilon^2}{144},\quad
 2\overline B_NR_\Lambda\leq\frac\epsilon2.
 \label{eq:half-budget}
\end{equation}
At $\epsilon=10^{-3}$, this requires the independent leakage-probability certificate $p\leq1/144000000$. All four readout cutoffs in Table~\ref{tab:cutoffs} still satisfy the halved truncation budget. Since $\sqrt{n_l}R_\Lambda+\delta\le5\epsilon/24<1$, the projected state is nonzero and can be normalized.

\subsection{Precision at fixed lattice size}
The product first falls below one at $j_0$. Once $j\ge1$ and $d_j>2A_g$, the denominator of the factor is $2c_N(j+1)^2[1-(s_a+A_g)/(c_N(j+1)^2)]$. Summing logarithms and using Stirling's formula gives
\begin{equation}
 \begin{aligned}[b]
 &-\log R_\Lambda=2\Lambda\log\Lambda\\
 &\quad{}-\big[2+\log(A_g/(2c_N))\big]\Lambda+O(\log\Lambda).
 \end{aligned}
 \label{eq:product-asymptotic}
\end{equation}
At fixed $\ell,g,E_*$ the remainder is uniform for $N\ge3$: $3/2<c_N\le81/52$, $0\le s_a\le2\ell E_*/(3g^2)$, and the logarithmic correction at each sufficiently large $j$ is uniformly $O((j+1)^{-2})$, hence summable. The linear coefficient describes the upper product, not necessarily the true vacuum tail. If $L=\log(2\overline B_N/\epsilon)\to\infty$, monotonicity of the product and comparison with $(1\pm\zeta)L/(2\log L)$ for fixed $\zeta>0$ give
\begin{equation}
 \Lambda_O=(1+o(1))\frac{L}{2\log L},\qquad
 L=\log\frac{2\overline B_N}{\epsilon}.
 \label{eq:precision-inversion}
\end{equation}
The one-step change of $-\log R_\Lambda$ is only $O(\log\Lambda)$, so integer rounding does not affect this equivalent. At fixed $N$ and $T$, the sufficient cutoff is therefore of order $\log(1/\epsilon)/\log\log(1/\epsilon)$. Equation~\eqref{eq:finite-cutoff}, rather than the asymptotic equivalent, determines a finite numerical budget.

At fixed lattice size, Eq.~\eqref{eq:true-tail} also establishes the cutoff required for uniform state approximation, complementing the sufficient certificate. Under the canonical subspace embedding, the maximal distance between a normalized vector in the spectral window and the hard-wall subspace is precisely $\|(1-P_\Lambda)P_*\|$. Enforcing normalization on the projected vector alters this distance by at most a factor of $\sqrt{2}$, provided the projection does not vanish. For the minimal integer cutoff achieving uniform state-vector error at most $\epsilon$, we therefore have
\begin{equation}
 \Lambda_{\min}(\epsilon)\sim
 \frac{\log(1/\epsilon)}{2\log\log(1/\epsilon)}
 \qquad(\epsilon\downarrow0).
 \label{eq:wall-necessary}
\end{equation}
Specifically, for any fixed $0<\zeta<1$ and all small enough $\epsilon$, the integer roundings of $(1\pm\zeta)$ times this asymptotic expression strictly bracket the threshold. This state-vector asymptotic does not determine the finite entries of Table~\ref{tab:cutoffs} or give a necessary cutoff for a particular readout or an arbitrary encoding. The volume-dependent lower bound also gives no necessity statement under lattice refinement.

A different cutoff scale controls energies at fixed lattice parameters. Measure the allowed absolute error in units of $\Delta_{\rm vac}$, and denote the tolerance in those units by $\epsilon$. Require every ordered level in the original window to satisfy it. The maximal shift is at least the vacuum shift and at most Eq.~\eqref{eq:spectral-accuracy-explicit}. Its logarithm in these units thus has the same leading term as Eq.~\eqref{eq:vacuum-energy-rate}. Inverting as for Eq.~\eqref{eq:wall-necessary}, the least cutoff meeting this spectral tolerance is asymptotic to $\log(1/\epsilon)/(4\log\log(1/\epsilon))$. This necessity comes from the absolute energy of the vacuum, not from an excitation frequency.

\subsection{Weak coupling at fixed evolution accuracy}\label{sec:weak-budget}

For a normalized initial state in the exact window, let $\Lambda_{\rm FT}$ and $\Lambda_c$ denote the first integer cutoffs at which $B_N q_\Lambda^{\rm FT}\le\epsilon$ and $B_N q_{c,\Lambda}\le\epsilon$, respectively. Either test certifies state-vector error at most $\epsilon$ for $|t|\le T$ when the wall evolution starts from the normalized projection. At fixed $N\ge3$, $\ell>0$, $E_*\ge0$, $T>0$ and fixed $0<\epsilon<1$, Appendix~\ref{app:weight} gives, as $g\to0$,

\[\frac{\Lambda_1}{\Lambda_{\rm FT}}\to2^{-1/4},\qquad \frac{\Lambda_{c_N}}{\Lambda_1}\to c_N^{-1/4}.\]

All three cutoffs grow as $g^{-1}\sqrt{\log(1/g)}$ up to constant factors. The product of the two ratios, $(2c_N)^{-1/4}$, compares explicit sufficient certificates, not minimal physical cutoffs. Table~\ref{tab:weights} separates the effect of changing the weight from the effect of the return flow. At $N=16$, the retained dimensions $2\Lambda+1$ are $307$, $265$ and $235$. Only the return-flow step crosses the $256$-dimensional register threshold in this row; at $N=8$ the weight change already suffices. Shell propagation alone also improves some budgets: at $g=0.30$ propagating the Fang--Tong v1 tail with $f_j$ reduces the cutoff from 48 to 45 for $N=8$ and from 51 to 47 for $N=16$. For each of the eight pairs $(N,g)$ with $N\in\{8,16\}$ and $g\in\{0.05,0.10,0.20,0.30\}$, the direct return-flow bound and the combined shell recursion certify the same integer cutoff for the trajectory test.

\begin{center}\begin{minipage}{\columnwidth}\centering\small\captionof{table}{First sufficient electric cutoff (independent binary qubits per link) for trajectory norm error $10^{-3}$, with $g=0.10$, $\ell=T=E_*=1$ and $N$ sites per direction. From left to right, the three budget columns use the Fang--Tong v1 bound, the modified weight construction at $c=1$, and the same construction with the Gauss return-flow coefficient $c=c_N$.}\label{tab:weights}
\begin{tabular}{rccc}\toprule
$N$ & Fang--Tong & $c=1$ & $c=c_N$\\ \midrule
8 & 146 (9) & 126 (8) & 112 (8)\\
16 & 153 (9) & 132 (9) & 117 (8)\\ \bottomrule
\end{tabular}\end{minipage}\end{center}

\begin{figure}[tbp]
 \centering
 \includegraphics[width=\linewidth,height=0.34\textheight,keepaspectratio]{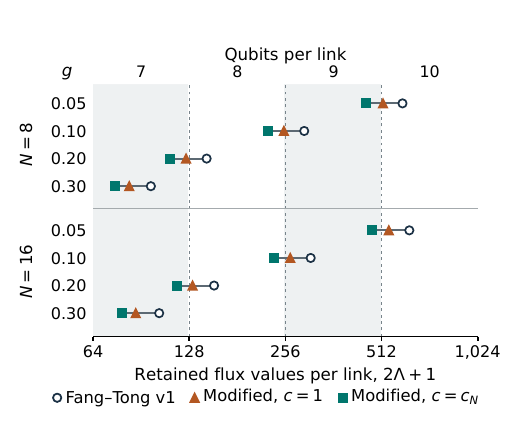}
 \caption{Sufficient link registers for a state-vector trajectory error of at most $10^{-3}$ at $\ell=T=E_*=1$. Each segment joins three certified dimensions $2\Lambda+1$ at fixed lattice size $N$ and coupling $g$: Fang--Tong v1 (open circle), the modified weight at $c=1$ (triangle), and the same construction at $c=c_N$ (square). Dashed lines mark powers of two on the logarithmic axis. A move to the left within a band reduces the number of flux values but leaves the binary budget, $\lceil\log_2(2\Lambda+1)\rceil$ qubits per link, unchanged.}
 \label{fig:register-thresholds}
\end{figure}

Relative to Fang--Tong v1, the modified weight at $c=1$ lowers the sufficient binary budget by one qubit per link in two of the eight cases in Fig.~\ref{fig:register-thresholds}. Four more cross a qubit threshold only when return flow is included. At $g=0.30$ the cutoff decreases at both steps on each lattice, but the bit count does not change.

For a different task, set $N=8$, $g=0.10$, $\ell=E_*=1$ and require absolute ordered-energy error at most $10^{-3}/\ell$. Using the same two-shell Ritz bound in every case, the product $R_j$, the published Fang--Tong tail, and the two modified weights certify cutoffs $283$, $113$, $100$ and $88$, respectively. The value $113$ uses their tail in our energy estimate. This certificate does not determine how many levels lie in the weak-coupling window. In Table~1(b), Fang and Tong~\cite{FangTong2026} compare their trajectory bound with the product bound in arXiv v2 of the present work~\cite{Liang2026v2}; for the trajectory comparison here we evaluate their full v1 action rather than an asymptotic approximation.

\subsection{Refinement at fixed coupling}
For fixed $\ell,g,E_*$ and $T>0$, one has $L=4\log N+\log(1/\epsilon)+O(1)$. In particular,
\begin{equation}
 \begin{aligned}
  &\epsilon_N=N^{-\alpha},\quad \alpha>0, \\
  &\Lambda_N=\left\lceil\beta\frac{\log N}{\log\log N}\right\rceil, \quad\beta>2+\frac\alpha2 \\
  &\Longrightarrow\quad 2\overline B_NR_{\Lambda_N}=o(N^{-\alpha}).
 \end{aligned}
 \label{eq:refinement-precision}
\end{equation}
Fixed nonzero tolerance recovers the sufficient coefficient $\beta>2$. The independent local register size is $O(\log\log N)$ bits per link, or $O(N^3\log\log N)$ bits over all links. At $T=0$, the certificate concerns static approximation of window states; its prefactor grows as $N^{3/2}$, giving the sufficient coefficient $\beta>3/4+\alpha/2$. At fixed $N,a,E_*$ the onset grows as $j_0\sim\sqrt{8/c_N}\,g^{-2}$ when $g\to0$. The fixed-coupling precision law is therefore not uniform at weak coupling.

For spectroscopy, under the same refinement at fixed $\ell$, $g>0$ and $E_*$, the volume factor is $J_an_p=3N^4/(g^2\ell)$. Along $\Lambda_N=\lceil\beta\log N/\log\log N\rceil$, the product is $R_{\Lambda_N}=N^{-2\beta+o(1)}$, and $R_{\Lambda_N-1}/R_{\Lambda_N}=O(\Lambda_N^2)=N^{o(1)}$. Hence, for $\beta>1$, $n_l R_{\Lambda_N}^2=N^{3-4\beta+o(1)}$ vanishes and the energy certificate is at most $N^{4-4\beta+o(1)}/\ell$ up to a fixed factor. Thus the hard-wall and untruncated ordered energies agree up to a vanishing absolute error, uniformly over the original window at fixed physical volume. This sufficient schedule does not use a gap bound uniform in $N$.

For a first-moment benchmark in the same sector use the fiber form inequality in Appendix~\ref{app:flow}. Evaluating this form on normalized states belonging to the spectral window and applying Eq.~\eqref{eq:vacuum-cost} yields
\begin{equation}
 \begin{aligned}
 \kappa_a c_N\langle E_l^2\rangle&\le E_*+mJ_a,\\
 \|Q_l(\Lambda)\psi\|&\le\frac{1}{\Lambda+1}\sqrt{\frac{A_g+s_a}{c_N}} .
 \end{aligned}
 \label{eq:energy-baseline}
\end{equation}
Here, the second relation expresses Markov's inequality applied to the link electric energy. Substituting this algebraic tail into the same bounded-readout precision budget leads to the sufficient first-moment cutoff
\begin{equation}
 \Lambda_{\rm moment}=\max\!\left\{0,
 \left\lceil\frac{2\overline B_N}{\epsilon}\sqrt{\frac{A_g+s_a}{c_N}}\right\rceil-1\right\}.
 \label{eq:energy-cutoff}
\end{equation}
At fixed physical parameters, the sufficient first-moment threshold in Eq.~\eqref{eq:energy-cutoff} grows as $O(N^4/\epsilon)$, whereas the spectral-window product yields the sub-logarithmic sufficient cutoff growth above. On the same spectral-window inputs, this comparison distinguishes the first-moment information used in Eq.~\eqref{eq:energy-baseline} from the stronger spectral information used by the tail product. Higher-moment and Monte Carlo approaches provide other routes to cutoff control~\cite{Arxiv2604_24896}.

\subsection{Matching regulators and error sources}
\label{subsec:regulator-comparison}

Specifying a cutoff requires defining both the subspace kept on each link and the operator action on that subspace. Under the hard-wall truncation adopted here, link shifts retain unit interior matrix elements while endpoint-crossing hops vanish. By contrast, cyclic digitizations wrap boundary hops to the opposite endpoint, whereas quantum-link formulations generally alter the interior matrix elements themselves~\cite{ChandrasekharanWiese1997}. Compressing the shifts of an enlarged cyclic register onto an interior flux interval reproduces the endpoint-zero hard-wall shifts~\cite{Haase2021}. Comparisons of regulator errors must match both retained spaces and operator actions. Further comparisons with finite-matrix, dual, and coupling-adapted schemes appear in Appendix~\ref{app:regulators}.

Local physical Hilbert-space truncation and entanglement truncation in tensor networks control distinct approximations. Tensor-network constructions can enforce Gauss's law while restricting virtual bond dimensions~\cite{BanulsCichy2020,ZoharBurrello2016PEPS}. Energy-cutoff Hamiltonian truncation instead selects an active subspace spanned by low-energy eigenstates of an unperturbed Hamiltonian, accounting for omitted states through effective interactions~\cite{CohenEtAl2022HTEffectiveTheory}. Here $P_*$ specifies physical initial states in the exact spectral subspace of the untruncated lattice theory, and $P_\Lambda$ imposes the local electric-flux cutoff.

Once $H_{a,\Lambda}$ is fixed, approximating its evolution introduces further errors. Product-formula and low-energy Trotter bounds require checking the encoded decomposition, cutoff-dependent local norms, and any additional state or energy-subspace hypotheses~\cite{ChildsEtAl2021Trotter,ChildsSu2019,SahinogluSomma2021,Arxiv2504_20746}. Locality decompositions and block encodings have their own spatial or oracle assumptions~\cite{HaahEtAl2023,LowChuang2019}. Gauge-invariant or stroboscopic circuits address time-discretized implementation, while improved Hamiltonians target spatial lattice artifacts~\cite{LammLawrenceYamauchi2019,BenderEtAl2018,CarenaEtAl2022}. Readout sampling and hardware noise are separate from the expectation-value error bounded here~\cite{KlcoRoggeroSavage2022,BauerEtAl2023Review}. A complete implementation budget combines the local-register bound with separately controlled preparation, evolution, and readout errors for compatible state, observable, time, and tolerance specifications.

%% file: sections/discussion.tex
\section{Discussion}\label{sec:discussion}

The interacting vacuum imposes a lower limit on a hard-wall representation even when $E_*=0$. A nonzero amplitude beyond every wall would establish that no finite cutoff is exact, but would say little about a target error. The quantitative lower bound comes from configurations with successively higher flux around the same plaquette. For absolute energies, the variational cost of missing part of the vacuum is at least $\Delta_{\rm vac}$ times its discarded probability. These two tests match the leading logarithms of the worst-case errors on the window: $-2\Lambda\log\Lambda$ for vectors and $-4\Lambda\log\Lambda$ for absolute energies in vacuum-gap units. They do not locate the error precisely at a finite wall. At $\Lambda=3$ in Sec.~\ref{sec:coherent-example}, for example, the lower and upper bounds on the same vacuum shift are separated by more than eleven orders of magnitude.

The factor of two in the leading cutoff is a distinction between tasks, not a factor of two in the number of qubits. The energy comparison uses small amplitudes on both sides of a boundary hop; uniform approximation of window vectors has to account for the exterior amplitude itself. Independent binary encoding takes $\lceil\log_2(2\Lambda+1)\rceil$ qubits per link. The same rounding is visible in Table~\ref{tab:cutoffs}: the energy criterion passes at a smaller cutoff than bounded readout at each lattice size, but both fit in six qubits per link. Excitation frequencies require another distinction. A frequency error is the difference of two nonnegative ordered energy shifts. Their common upper bound controls the difference, but the shifts can cancel. The necessity established by the absolute vacuum energy therefore does not establish a necessary cutoff for an individual excitation frequency.

Finite-matrix data can also certify properties of the untruncated lattice spectrum. Suppose the ordered energies of the finite wall have certified lower and upper endpoints $\underline\lambda_k\le E_k^\Lambda\le\overline\lambda_k$, in the same sector and at a cutoff where both truncation certificates apply. Subtract the vacuum-only truncation bound $\epsilon_{\rm vac}$ from $\underline\lambda_0$ to get a lower bound on the original $E_0$. If $\overline\lambda_k\le\underline\lambda_0-\epsilon_{\rm vac}+E_*$, min--max places the original level indexed by $k$ inside the window. Only then do we use the full-window truncation bound $\varepsilon_\Lambda$: the absolute energy lies between $\underline\lambda_k-\varepsilon_\Lambda$ and $\overline\lambda_k$. Rigorous counts at the two shifted thresholds of Corollary~\ref{cor:finite-spectrum} determine the dimension if they agree; a level near the edge can leave the count undecided. This test concerns ordered energies, not the spectral support of finite-wall eigenvectors. In the strong-coupling example an analytic band argument, not a finite-matrix count, identifies the vacuum and all 162 states of the first excited band.

For real-time evolution, the comparison depends on what is known about the input. Projecting a state and dividing by its surviving norm is a conditional operation; it does not generally define a deterministic encoding channel for an unknown state. If the wall projection is injective on the window, its polar isometry acts linearly on the entire subspace. The bound on the unhalved diamond norm of the difference between the encoded and exact evolution channels covers inputs entangled with an arbitrary reference. It does not supply a circuit for the isometry. For a state or logical subspace prepared inside the wall, both evolutions can start from the same input, with leakage outside the original window budgeted independently. The two four-link pulses in Sec.~\ref{sec:coherent-example} prepare one prescribed state. Table~\ref{tab:trial}, by contrast, compares two channel bounds for the same trial qubit: exploiting its known flux support improves on the window-plus-leakage estimate at all three durations.

A small probability at high flux can dominate $\langle E_l^2\rangle$ even when two state vectors are close. For an exact-window input, Sec.~\ref{sec:electric-square} bounds the expectation difference inside the wall using the norm of the clipped operator and the trajectory error. A weighted sum of discarded-flux probabilities controls the part outside. For inputs with out-of-window preparation leakage, the leakage norm alone does not supply the required moment bound.

The size of a sufficient binary register can change in a discrete step, even when the change in the tail bound is smooth. To see the role of the electric constraints, fix flux $k$ on one link. Among real currents with zero divergence and zero global flux, the least total squared flux is $c_N k^2$, where $c_N=3N^3/[2(N^3-1)]$. The other links must therefore carry at least $(c_N-1)k^2$ in squared flux. Integer currents obey this bound without necessarily attaining the real minimum. Table~\ref{tab:weights} fixes $g=0.10$, $\ell=T=E_*=1$ and a trajectory vector-norm tolerance of $10^{-3}$. At $N=16$, the Fang--Tong certificate and the modified weight alone select 307 and 265 retained flux values, respectively; both fit in nine qubits per link under independent binary encoding. Adjusting the parameter from $1$ to $c_N$ in this construction lowers the certified dimension from 265 to 235, below the eight-qubit threshold of 256 flux values. At $N=8$ the weight change alone crosses the same threshold. This is a comparison of sufficient registers, not a hardware lower bound.

Increasing precision, decreasing coupling and refining the spatial mesh are different limits. At fixed physical box size, positive bare coupling, window width and duration, the refinement schedules make the expectation error vanish for uniformly bounded families of observables. This compares evolution of exact-window inputs with evolution from their normalized wall projections, using the observables compressed to the wall. If the untruncated expectations already have a continuum limit, their hard-wall counterparts have the same limit. Following a running bare coupling instead would require estimates on the window and its tail along that trajectory; fixed-coupling transfer does not construct a continuum limit. Even on one lattice the high-precision converse does not determine necessity as $g\to0$. The first shell where the fixed-coupling product contracts moves out as $g^{-2}$, while the weighted estimate certifies a cutoff of order $g^{-1}\sqrt{\log(1/g)}$ at fixed positive duration and tolerance. To decide whether low-energy evolution actually requires that flux range, one needs a converse whose constants remain controlled as the coupling decreases.

%% file: data-availability.tex
\paragraph*{Data availability.} The calculation files on \mbox{Zenodo}~\cite{LiangQianZenodo2026} reproduce the reported cutoff budgets, ordered-energy and contrast bounds, plaquette counts, and trial-channel comparison.

%% file: sections/auxiliary.tex
\section{Mean energy, the Weyl probe, and compensating flux}
\label{app:auxiliary}
\subsection{Pure-state mean-energy counterexample}

For $E_*>0$, a pure-state counterexample is obtained by orthogonalizing and normalizing $|e_M\rangle$ against $|\Omega\rangle$ to form $|\eta_M\rangle$, and constructing the coherent superposition $\sqrt{1-\tilde w_M}|\Omega\rangle+\sqrt{\tilde w_M}e^{i\theta}|\eta_M\rangle$ with weight $\tilde w_M$ tuned to maintain mean excitation $E_*$. Because $|\eta_M\rangle$ is orthogonal to the vacuum eigenstate, the cross term in the energy expectation vanishes identically for all phases $\theta$. The removed vacuum overlap tends to zero as $M\to\infty$, so $\tilde w_M$ is of order $M^{-2}$ and $\|Q_l(M-1)|\eta_M\rangle\|^2\to1$. Selecting $\theta$ so that the high-flux cross term is non-negative therefore gives a tail at least of order $M^{-2}$.

\subsection{Nontriviality of the Weyl probe}

Because electric flux has an integer spectrum, having a nonvanishing continuum curl does not by itself prevent all lattice circulation phases from becoming integer multiples of $2\pi$. The amplitude restriction $0<g\ell|\alpha_f|<3\pi$ guarantees the bound below on every elementary plaquette $p$: with $f_l=f_i(ax)$ evaluated on each positively directed link $l=(x,i)$, the discrete circulation $\phi_p=ga\sum_{l\in\partial p}c_{pl}f_l$ obeys $|\phi_p|\leq2g\ell|\alpha_f|/N<2\pi$ for all $N\geq3$. For an elementary $(1,2)$ plaquette adjacent to $x_2=0$, this circulation evaluates to $\pm ga\alpha_f\sin(2\pi/N)$, which is strictly nonzero. An elementary unit flux loop on that plaquette defines a physical state that possesses an eigenvalue under $W_f(a)$ distinct from that of the trivial zero-electric state.

\subsection{Energy of the compensating flux}
\label{app:flow}

Let $Z$ denote the real vector space of link currents having zero lattice divergence and vanishing total flux across each of the three coordinate planes. On a connected cubic lattice of $N^3$ sites, the node-link incidence matrix has rank $N^3-1$. The three coordinate-plane cut functionals are linearly independent on the kernel of the incidence matrix, as demonstrated by considering straight loops winding around each torus cycle; hence $\dim Z=2(N^3-1)$. Under lattice translations and cubic point-group symmetries, the orthogonal projector $\Pi$ onto $Z$ has constant diagonal elements, whose trace yields $\Pi_{ll}=1/c_N$. For a prescribed link flux $k$ on link $l$, the unique flow in $Z$ of minimal Euclidean norm is $z_*=k\Pi e_l/\Pi_{ll}$, where $e_l$ is the unit basis vector on link $l$. Indeed, the difference between any two admissible extensions is orthogonal to $\Pi e_l$. Writing $t$ for the environment components of $z_*$, every allowed integer-valued environment configuration $n$ satisfies the orthogonal decomposition
\begin{equation}
 \begin{aligned}
 \|t\|^2&=(c_N-1)k^2,\\
 \|n\|^2&=\|n-t\|^2+(c_N-1)k^2 .
 \end{aligned}
 \label{eq:fiber-geometry}
\end{equation}

Consider a fiber state represented by an environment angle wavefunction $u$. Because the Fourier expansion of $u$ is supported on such integer labels $n$, its kinetic form is the sum of $\kappa_a(c_N-1)k^2\|u\|^2$ and the covariant gradient form $\kappa_a\|(-i\nabla-t)u\|^2$. The potential in $H_{\rm env}$ acts as a real multiplication operator. From the polar representation of $u$ away from its zeros and standard quadratic-form regularization, one obtains the pointwise diamagnetic inequality $|\nabla |u||\le |(-i\nabla-t)u|$. The shifted environment form evaluated on $u$ is therefore bounded below by the unshifted form evaluated on $|u|$, which is periodic and has the same norm. Because $e_{\rm env}$ is the ground-state infimum of the unshifted periodic environment Hamiltonian on the entire torus, the energy of $|u|$ is at least $e_{\rm env}\|u\|^2$, without imposing any Fourier-support constraint on $|u|$. This verifies the claimed fiber inequality without requiring $t$ to be an integer gauge vector. The vacuum comparison in Eq.~\eqref{eq:vacuum-cost} relies on the same value $e_{\rm env}$ because the positive environment ground state is invariant under the restricted gauge and flux translations: tensored with $|E_l=0\rangle$, it is an admissible neutral trial state. Summing these form bounds across all fibers and reintroducing the nonnegative incident plaquette terms establishes the operator form inequality $H_a\ge e_{\rm env}+\kappa_a c_N E_l^2$ throughout the neutral sector.

%% file: sections/coherent-proof.tex
\section{Spectral and readout certificate for the coherent example}
\label{app:coherent-proof}

The following estimates refer to the original physical neutral sector at the parameters of Sec.~\ref{sec:coherent-example}. Write
\begin{equation}
 \begin{aligned}
  &H_a=\nu+h,\qquad h=K+W, \\
  &K=54\sum_lE_l^2,\quad W=-\frac1{12}\sum_pC_p, \\
  &\nu=\frac{27}{4},\quad\|W\|\leq\nu.
 \end{aligned}
 \label{eq:example-split}
\end{equation}
Gauss's law makes the flux through parallel coordinate cuts equal. Zero global flux thus implies that $\sum_x n_{x,i}=3 F_i(n)$ vanishes for any divergence-free electric configuration $n$ in the neutral sector, with $F_i(n)$ the eigenvalue of the cut-flux operator $F_i$. Since $n^2\equiv n\pmod2$, its kinetic quadratic form $\sum_{x,i}n_{x,i}^2$ is even. Value $2$ would require two unit edges to form a conserved flow, impossible on the simple cubic torus. Value $4$ consists of four unit edges forming an elementary square: a lone magnitude-two edge violates Gauss's law, and a length-four cycle cannot wind on the $N=3$ torus. Hence the first kinetic eigenspace consists of the $2n_p=162$ oriented unit plaquettes. All other nonzero levels are at least $6\kappa_a=324$. In particular, the length-three wrapping loops are excluded by the global-flux constraint.

The operator inequalities $K-\nu\leq h\leq K+\nu$ and the min--max principle bound every ordered eigenvalue of $h$ by the corresponding kinetic eigenvalue plus or minus $\nu$. Its compact resolvent follows from the bounded perturbation. There is one ground eigenvalue $e_0=E_0-\nu\in[-\nu,0]$, exactly $162$ eigenvalues, counted with multiplicity, in $[837/4,891/4]$, and all higher eigenvalues are at least $1269/4$. The ground upper bound follows by evaluating $h$ in the zero-electric state. These intervals are disjoint; no simplicity assumption is made within the first excited band.

A sharper vacuum estimate controls the plaquette transition. Let $Q_{\rm bare}=I-|0\rangle\langle0|$ and $\Omega=\alpha|0\rangle+q$ with $\alpha>0$ and $q\perp|0\rangle$. The ground equation gives
\begin{equation}
 \begin{aligned}
  &(Q_{\rm bare}hQ_{\rm bare}-e_0)q=-\alpha W|0\rangle, \\
  &Q_{\rm bare}hQ_{\rm bare}-e_0\geq\frac{837}{4}Q_{\rm bare}, \\
  &\|W|0\rangle\|^2=\frac9{32}.
 \end{aligned}
 \label{eq:example-vacuum}
\end{equation}
The last identity counts the orthogonal oriented plaquette states. It implies $\|q\|\leq\alpha/(279\sqrt2)$ and $\|\Omega-|0\rangle\|\leq1/279$. Projecting the same eigenvalue equation onto $|0\rangle$ further gives $-1/744\leq e_0\leq0$. The first-band excitation frequencies therefore lie in $[837/4,891/4+1/744]$, whereas all higher excitations are at least $1269/4$. Thus the exact window of width $270$ contains precisely the vacuum and this band.

These estimates also give a lower certificate for the vacuum-energy shift at $\Lambda=3$. The excitation interval gives $\Delta_{\rm vac}\ge 837/4$, while $\|q\|\le\alpha/(279\sqrt{2})$ implies $c_0^2=\alpha^2\ge155682/155683$. The circulation $4\partial p$ lies outside the wall. Equations~\eqref{eq:vacuum-recursion} and~\eqref{eq:vacuum-energy-lower} therefore bound the vacuum shift from below by $(837/4)(155682/155683)(1/5346)^8/(4!)^4$, approximately $9.45\times10^{-34}$ in the energy units used here.

For $c=C_p|0\rangle$, one has $\|c\|=1/\sqrt2$, $Kc=216c$, and $(h-216)c=Wc$. The spectrum of $h$ outside the first band is at distance at least $405/4$ from $216$. Its complementary resolvent consequently gives
\begin{equation}
 \begin{aligned}
  &\|(I-P_{\rm band})c\|\leq\frac1{15\sqrt2}, \\
  &\|P_{\rm band}c\|\geq\frac{4\sqrt7}{15}, \\
  &\|P_{\rm band}C_p\Omega\|\geq\frac{4\sqrt7}{15}-\frac1{279}.
 \end{aligned}
 \label{eq:example-transition}
\end{equation}
This proves the transition bound in Eq.~\eqref{eq:coherent-wavepacket} for the ordinary plaquette operator.

The vacuum--wavepacket interference has a positive spectral measure. For the state in Eq.~\eqref{eq:coherent-wavepacket},
\begin{equation}
 \begin{aligned}
  &M(t)=\frac12\langle\Omega,C_p\Omega\rangle \\
  &\quad{}+\frac12\langle U(t)\chi,C_pU(t)\chi\rangle \\
  &\quad{}+\frac1z\int\cos(\omega t)\,d\mu(\omega), \\
  &\mu(\mathbb R)=z^2.
 \end{aligned}
 \label{eq:example-signal-measure}
\end{equation}
Here $\mu$ is the spectral measure of $P_{\rm band}C_p\Omega$ for $H_a-E_0$, supported in the certified excitation interval. At $t_0=\pi/216$, every phase differs from $\pi$ by at most
\begin{equation}
 d=\left(\frac{27}{4}+\frac1{744}\right)\frac\pi{216}<\frac1{10}.
 \label{eq:example-phase}
\end{equation}
Moreover, $\|(h-216)P_{\rm band}\|\leq\nu$ bounds the change in the excited-sector expectation term of Eq.~\eqref{eq:example-signal-measure}, including its factor one half, by $\nu t_0=\pi/32$ after removal of the common band-center phase. It follows that
\begin{equation}
 \begin{aligned}
  &M(0)-M(t_0)\geq z(1+\cos d)-\frac\pi{32} \\
  &>\frac{701}{1000}\frac{399}{200}-\frac{11}{112}>1.30.
 \end{aligned}
 \label{eq:example-drop}
\end{equation}
The rational bounds use $\pi<22/7$ and $\cos d\geq1-d^2/2$. Finally, applying Eq.~\eqref{eq:observable-error} at both endpoints with Eq.~\eqref{eq:coherent-budget} gives the truncated change exceeding $1.29$. Every numerical decision in this certificate can be reproduced with integer and rational arithmetic. The spectral window was identified analytically for the untruncated Hamiltonian.

\subsection{Leakage and hard-wall errors for the trial qubit}
\label{app:trial}
Energy scales in the residual bounds are measured with respect to $H_a$, including the additive magnetic constant $\nu=27/4$. The orthogonal complement to the 163-dimensional window begins at or above $1269/4+\nu=324$. Using the definitions in Eq.~\eqref{eq:example-split}, unperturbed energies under $\nu+K$ are $\nu$ for $|0\rangle$ and $\nu+216=891/4$ for $\chi_0$. Acting on $|0\rangle$, $W$ generates 162 distinct oriented plaquette excitations, each with amplitude $-1/24$. Acting on the two branches of $\chi_0$ produces 324 elementary hopping terms before collecting configurations: the vacuum appears twice, while 322 distinct configurations appear with multiplicity one. Distinct plaquettes intersect on at most one link, so the difference between two boundary loops matches $2\partial p$ only when both loops lie on $p$. None of these resulting states matches the flux configuration of an isolated square. Consequently, $W|0\rangle$ and $W\chi_0$ are mutually orthogonal, with squared norms $9/32$ and $163/576$. For either of these two unperturbed eigenvectors $x$ with energy $e$, the spectral theorem bounds $\|(1-P_*)x\|\le\|Wx\|/(324-e)$. Weighting both column bounds by their respective amplitudes in $\psi_{\rm prep}$ yields
\begin{equation}
 \delta_{\rm prep}\le 1/846+\sqrt{326}/4860<0.004898 .
 \label{eq:prep-leak}
\end{equation}
The projected operator $A=J^\dagger H_a J$ contains these unperturbed diagonal energies and an off-diagonal coupling $-\sqrt2/24$. Bounding the shift of the larger eigenvalue by the squared off-diagonal coupling divided by 216 ensures $A\le bI$ with $b=891/4+1/62208$. Projecting out the subspace $\operatorname{Ran}J$ leaves the residual Gram matrix $\mathcal R^\dagger\mathcal R=\operatorname{diag}(160,161)/576$. The residual bound from Sec.~\ref{sec:dynamics} then establishes $\delta_J\le\sqrt{161}/[24(324-b)]<0.005222$.

Under unperturbed evolution, the expectation value of $C_p$ along $\psi_{\rm prep}$ evaluates to $\cos(216t)/\sqrt2$. Due to the orthogonality of the two hopping states, $\|W e^{-it(\nu+K)}\psi_{\rm prep}\|^2=325/1152$ holds for all $t$. Duhamel's formula bounds the deviation from unperturbed evolution by $|t|\sqrt{325/1152}$ in the full lattice theory as well as in any truncation wall enclosing the unperturbed path. Initial expectation values match, guaranteeing that the dynamical contrast at $t_0=\pi/216$ reaches at least $\sqrt2-2t_0\sqrt{325/1152}>1.3987$ in either setting. Alternatively, inserting the unrounded bound from Eq.~\eqref{eq:prep-leak} into Eq.~\eqref{eq:trial-channel} at cutoff $\Lambda=3$ recovers the measurement error certificate and transferred contrast stated in the text.

To evaluate the final column of Table~\ref{tab:trial}, consider evolution inside the truncation wall $\Lambda=3$ generated by the shifted Hamiltonian $K+W_\Lambda$, where $W_\Lambda=P_\Lambda WP_\Lambda$ denotes the projected interaction. Let $P_{m,l}=\mathbf1_{\{|E_l|\ge m\}}$ and define $Y_m(f)^2=\sum_l\|P_{m,l}f\|^2$. For any link, the incoming transition block mapping $|E_l|=m-1$ to $|E_l|\ge m$ (for $m=2,3,4$) has norm bounded by $v=4J_a/2=1/6$. Each incident plaquette contributes only its raising hop on the positive side and its lowering hop on the negative side. The two sign blocks have orthogonal input and output spaces; at $m=4$ the hop is the uncompressed interaction across the wall. Partitioning the exterior space by the first violating link establishes that the outward coupling norm is bounded by $vY_3(f)$, fully accounting for states exiting across multiple links simultaneously.

For $f(t)=e^{-it(K+W_\Lambda)}Jx$ with normalized input $\|x\|=1$, the higher flux sectors $m=2$ and $m=3$ initially contain zero amplitude. Applying variation of constants to each projected sector and invoking the triangle inequality across the direct sum yields $Y_m(f(t))\le v\int_0^{|t|}Y_{m-1}(f(\operatorname{sgn}(t)s))\,ds$. Since $\sum_lP_{1,l}\le K/54$, energy conservation combined with $\|W_\Lambda\|\le27/4$ guarantees $Y_1\le\sqrt{17}/2$. Integrating twice in time and once across the boundary bounds the state isometry error by $v^3|t|^3\sqrt{17}/12$. Doubling this value bounds the channel error by $\sqrt{17}|t|^3/1296$.

A linear-in-time bound is obtained by evaluating each basis column of the trial space separately. For either column vector $j$ satisfying $Kj=e j$ (with $e=0$ or $216$), let $f=e^{-it(K+W_\Lambda)}j$ and $w=W_\Lambda j$, which satisfy $(K+W_\Lambda-e)f=e^{-it(K+W_\Lambda)}w$. Here $\|w\|\le3/4$, and $K/54\le16$ on its support. For any state $z$ initially supported on $K/54\le k$, energy conservation implies $Y_m(e^{-it(K+W_\Lambda)}z)\le\sqrt{k+1/4}\,\|z\|/m$. When $|E_l|\ge3$, Gauss's law requires at least three units of compensating absolute flux at each endpoint of $l$, on disjoint sets of other links. Integer flux quantization therefore imposes $K/54\ge15$. The compression of $K+W_\Lambda-e$ to this tail is bounded below by $\gamma I$, where $\gamma=15\cdot54-216-27/4=2349/4$. Projecting the residual equation onto this tail, inverting the block, and applying the transition bound $v$ yields
\begin{equation}
 Y_3(f)\le\gamma^{-1}\left[
 \frac{3}{4}\frac{\sqrt{16+1/4}}{3}
 +v\frac{\sqrt{4+1/4}}{2}\right].
 \label{eq:trial-exit}
\end{equation}
Multiplying the right-hand side by $vT$ accounts for the integrated boundary flux, scaling by $\sqrt2$ combines both basis columns in operator norm, and multiplying by two converts the bound to channel diamond distance. The resulting linear coefficient of $T$ is strictly less than $0.00095$. The pointwise minimum of the cubic, linear, and trivial channel bounds produces the trial-specific column. Both channel bounds compare the same full and naturally included hard-wall outputs, and hold for arbitrary entanglement of the trial qubit with a reference.

%% file: sections/regulator-details.tex
\section{Additional regulator comparisons}
\label{app:regulators}

Finite-matrix constructions, quantum link models, and D-theory realize exact continuous gauge symmetries with finite local operator algebras~\cite{Horn1981,OrlandRohrlich1990,ChandrasekharanWiese1997,Wiese2013,BrowerEtAl1999,BrowerEtAl2004DTheory}. Discrete-subgroup digitizations preserve subgroup multiplication but introduce discretization artifacts~\cite{AlexandruEtAl2019,AlamEtAl2022}; group decimation and representation subduction analyze and mitigate such artifacts in effective actions~\cite{JiLammZhu2020,AssiLamm2024}. Proposed fuzzy regulators invoke universality arguments to recover continuous gauge fields~\cite{AlexandruEtAl2024Fuzzy}, while orbifold formulations of $U(k)$ gauge theory use noncompact matrix variables whose oscillator encodings require auxiliary cutoffs~\cite{BuserEtAl2021}.

Variable choices reorganize physical constraints. Dual descriptions reduce independent variables in two spatial dimensions but retain local constraints in three, including in formulations with matter~\cite{KaplanStryker2020,BenderZohar2020}. Cyclic formulations separate group discretization from the retained flux range~\cite{Haase2021}, while coupling-adapted magnetic grids target configurations dominant at weak coupling~\cite{BauerGrabowska2023}. For $\mathrm{SU}(2)$, prepotentials retain gauge and link-matching constraints~\cite{Mathur2005}; loop-string-hadron variables internalize non-Abelian Gauss laws with residual Abelian matching~\cite{RaychowdhuryStryker2020}, with spectral and dynamical comparisons in $1+1$ dimensions~\cite{DavoudiRaychowdhuryShaw2021}. Plaquette-Krylov methods with large-$N_c$ terms~\cite{CiavarellaEtAl2025Efficient} and site-singlet irreducible-representation cutoffs with approximate preparation~\cite{PerturbationIrrep2026} have small-system tests. Compact-$U(1)$ basis comparisons show that efficiency is regime-dependent~\cite{JakobsEtAl2026TruncatedBases}.

Scalar-field algorithms distinguish spatial discretization from field digitization and use energy bounds to control state truncation~\cite{JordanLeePreskill2012,JordanLeePreskill2014}; oscillator bases and field-coordinate grids tailor representations to low-lying modes~\cite{Somma2016,MacridinEtAl2018,KlcoSavage2019}. Gauge-invariant tensor networks~\cite{BanulsCichy2020,TensorNetworksBeyond2025} embed symmetry into local tensors for quantum links or continuous groups~\cite{RicoEtAl2014,TagliacozzoCeliLewenstein2014,ZoharBurrello2016PEPS}. Applications include Schwinger-model ground states and a proposed $(2+1)$-dimensional projected entangled pair state ansatz for quantum electrodynamics (QED)~\cite{ZappOrus2017}, and higher-dimensional quantum-link systems with matter~\cite{FelserEtAl2020,MagnificoEtAl2021}. Periodic Gaussian variational states treat compact QED without an explicit flux cutoff, but approximate infinite sums and variational dynamics~\cite{BenderEtAl2020VMC}.

Energy-cutoff Hamiltonian truncation compensates omitted unperturbed-energy states through renormalization or effective operators~\cite{RychkovVitale2015HamiltonianTruncation,EliasMiroRychkovVitale2017NLO,CohenEtAl2022HTEffectiveTheory}. Ultraviolet-divergent cases require counterterms~\cite{EliasMiroIngoldby2023Counterterms,DeloucheEliasMiroIngoldby2024HT}, as do light-front momentum-Fock treatments of QED~\cite{DraperHidalgoIlderton2026}. The distinctions needed to combine such approximations with the present local-flux bound are summarized in Sec.~\ref{subsec:regulator-comparison}.

%% file: sections/weighted-proof.tex
\section{Weighted localization and cutoff comparison}\label{app:weight}

Introduce the rescaled energies $e_{\rm sc}=a\tau E_0$ and $\delta_{\rm sc}=a\tau E_*$. Ground-state positivity and symmetry guarantee that $E_0$ coincides with the vacuum energy featured in the local comparison. Discarding the positive semidefinite incident magnetic terms and invoking the fiberwise inequality of Appendix~\ref{app:flow} leads to $\mathcal K-e_{\rm sc}\ge c\tau^2 E_l^2/2-4\tau$ when $c=c_N$, and consequently for $c=1$ as well. Setting $(z)_+=\max(z,0)$, we define for each $n\in\mathbb Z$ and truncation cap $b_0>0$

\begin{equation*}\begin{aligned}\phi(n)&=\frac{\mathcal I_8(\sqrt c\,\tau(|n|-1)_+)}{\sqrt c\,\tau},\\ \phi_{b_0}(n)&=\min\{b_0,\phi(n)\},\qquad W=e^{\phi_{b_0}(E_l)}.\end{aligned}\end{equation*}

The bounded weight is diagonal in the electric basis and commutes with the charge, winding and kinetic operators, so it preserves the physical sector and the kinetic domain. At $n=0$ both neighboring differences vanish; for either neighbor at any other $n$, monotonicity of $\mathcal I_8^\prime$ bounds the absolute difference of the capped weights in the exponent by $\operatorname{arcosh}(1+c\tau^2 n^2/16)$. In the discrete weighted-form estimate~\cite[Eq.~(38)]{FangTong2026}, each of four incident plaquettes has a hopping coefficient $-1/2$ in each direction. The correction at $n$ is bounded by twice the sum over the two neighbors of $\cosh(\phi_{b_0}(n\pm1)-\phi_{b_0}(n))-1$, hence by $c\tau^2n^2/4$.

Let $\psi$ be a normalized state within the finite-dimensional spectral window that attains the maximum of $\|W\psi\|$, and set $v=W\psi$ along with its flux components $v_n=\mathbf1_{\{E_l=n\}}v$. It follows that $P_*W^2P_*\psi=\|v\|^2\psi$. Window invariance implies the inequality $\operatorname{Re}\langle W^2\psi,(\mathcal K-e_{\rm sc})\psi\rangle\le\delta_{\rm sc}\|v\|^2$. Combining this with the weighted-form lower bound gives

\[\sum_{n\in\mathbb Z}(c\tau^2 n^2/4-4\tau-\delta_{\rm sc})\|v_n\|^2\le0.\]

Define $r_0>0$ through $r_0^2=8(4\tau+\delta_{\rm sc})/(c\tau^2)$. For configurations with $|n|\ge r_0$, the bracketed factor is bounded below by $4\tau+\delta_{\rm sc}$, whereas on the domain $|n|<r_0$ it is bounded below by $-(4\tau+\delta_{\rm sc})$. As a result, the weighted mass lying outside this radius cannot exceed the mass contained inside. Taking into account the property $\mathcal I_8(y)\le y^2/(4\sqrt2)$ together with $\lceil r_0\rceil-1\le r_0$, the exponent in the interior region is bounded above by $\sqrt{2/c}\,A_{\rm w}$. This furnishes a cap-independent upper bound on $\|WP_*\|$ of $\sqrt2\exp(\sqrt{2/c}\,A_{\rm w})$. For $|n|>j$, the uncapped exponent is at least $G_c(j)$. Choosing $b_0\ge G_c(j)$ therefore gives $\|Q_l(j)P_*\|\le e^{-G_c(j)}\|WP_*\|\le q_{c,j}$, as asserted in Proposition~\ref{prop:weighted}.

To establish the pointwise inequality, treat $j$ as a continuous variable on $[0,\infty)$ and define $x=\tau j/4$. Taking derivatives of both action functions yields $F^\prime(j)=\operatorname{arsinh}x$ alongside $G_c^\prime(j)=2\operatorname{arsinh}(\sqrt{c/2}\,x)$. Whenever $1\le c\le2$, the concavity of $\operatorname{arsinh}$ on $\mathbb R_{\ge0}$ establishes that $G_c\ge\sqrt{2c}\,F$. Under the assumption that $q_j^{\rm FT}\le1$, the action satisfies $F\ge A_{\rm w}+\log\sqrt2$. Using these two inequalities, including the difference of prefactors, gives

\[\begin{aligned}\log(q_{c,j}/q_j^{\rm FT})\le{}&(\sqrt{2/c}-\sqrt{2c})A_{\rm w}\\ &-(\sqrt{2c}-1)\log\sqrt2<0.\end{aligned}\]

The comparison holds for $c_N$ because $3/2<c_N\le81/52$. At fixed $j$ the action $G_c(j)$ is nondecreasing in $c$ by the derivative formula above, and its exponential prefactor in $q_{c,j}$ decreases. At fixed $\tau>0$ the large-argument law $\mathcal I_8(y)=2y\log y+O(y)$ recovers the leading $-2j\log j$ in the logarithm of the new direct bound. This is an upper estimate; the independent vacuum argument in Proposition~\ref{prop:sharp} supplies necessity.

For weak coupling take $N\ge3$, $\ell>0$, $E_*\ge0$, $T>0$ and the state-vector tolerance $0<\epsilon<1$ fixed. With $B_N$ as in Eq.~\eqref{eq:trajectory}, let $z=\log(\sqrt2 B_N/\epsilon)$. Then $z\to\infty$ and $\tau z\to0$ as $g\to0$. Let $j_{\rm FT}$ and $j_c$ solve $F(j_{\rm FT})=z+A_{\rm w}$ and $G_c(j_c)=z+\sqrt{2/c}\,A_{\rm w}$, respectively. Inverting near zero using $\mathcal I_b(y)=y^2/(2\sqrt b)+O(y^4)$ gives $\tau j_{\rm FT}^2/8\sim z$ and $\sqrt c\,\tau j_c^2/(4\sqrt2)\sim z$, respectively. Both thresholds diverge, so integer rounding leaves the ratio unchanged.